\documentclass[journal]{IEEEtran}
\usepackage{cite}
\usepackage[fleqn]{amsmath}
\usepackage{amsfonts,amssymb}
\usepackage{newtxtext}
\usepackage[varg]{newtxmath}
\usepackage{graphicx}
\usepackage{subcaption}
\usepackage{booktabs}
\usepackage{textcomp}
\usepackage{array}
 \usepackage{multirow}
 
\usepackage[ruled,vlined,noend]{algorithm2e}
\usepackage[
  setpagesize=false,
  colorlinks=true,
  linkcolor=blue,
  citecolor=blue,
  urlcolor=blue
]{hyperref}

\title{Channel Modeling and LED Spot Detection for \\Dense Image-Sensor Visible Light Communication}

\author{Tianhao Shi,~\IEEEmembership{Student Member,~IEEE,} Shan Lu,~\IEEEmembership{Member,~IEEE,} and~Takaya Yamazato,~\IEEEmembership{Senior Member,~IEEE}
 
\thanks{This work was partly supported by the JSPS KAKENHI Grant Number 25K00371.}
\thanks{T. Shi, S. Lu, and T. Yamazato are with Nagoya University, Nagoya, Japan (e-mail: shi.tianhao.u1@s.mail.nagoya-u.ac.jp; shan.lu.jp@ieee.org; yamazato@ieee.org).}%
}

\begin{document}
\maketitle

\begin{abstract}
High-density LED arrays enable high-speed transmission in image-sensor-based visible-light communication (VLC) systems. However, when optical spots become blurred and spatially overlapped due to focal shift, resolution limitations, or interference, severe inter-symbol interference (ISI) occurs, significantly degrading decoding performance. \textbf{Furthermore, radial distortion introduces geometric deformation of the LED grid, while vignetting leads to incomplete and asymmetric spot shapes at the periphery, both of which further hinder reliable signal detection.} Existing methods mitigate ISI by reducing LED transmission signaling density.

This paper proposes a robust decoding framework that maintains full LED signaling density. We introduce a pilot-aided geometric recognition method that uses a PSF-constrained Hough transform and circle-center alignment refinement. \textbf{In addition, radial distortion correction and vignetting-aware compensation are incorporated to restore geometric consistency and suppress edge-related detection errors.} By leveraging prior structural knowledge from pilot frames, the system effectively separates overlapping LED signals under severe optical distortion.

Experimental results on a real-world VLC testbed confirm that the proposed method achieves superior decoding accuracy and throughput compared to conventional Hough-based and low-density baseline methods. The results highlight its potential for high-efficiency VLC applications in interference-prone environments.
\end{abstract}

\begin{IEEEkeywords}
Image sensor communication, Visible Light Communication, Blur LED, LED Array, Hough transform
\end{IEEEkeywords}

\section{Introduction}
Visible Light Communication (VLC) is an emerging technology that utilizes visible light for data transmission \cite{vlc2}\cite{vlc}.
Leveraging light-emitting diodes (LEDs) for both illumination and data transmission, VLC systems are particularly attractive for smart homes, vehicle-to-vehicle communication, and Internet of Things (IoT) applications.

Image sensor communication (ISC) \cite{Image_sensor_communication} is a type of VLC technology in which LEDs transmit signals, and the receiver is an image sensor.
In the practical ISC system, the optical signals emitted from multiple LEDs are captured by CMOS image sensors as circular light spots, such as, traffic signals can function as transmitters to convey information, while in-vehicle cameras act as receivers to interpret these signals, laying the groundwork for future intelligent driving applications\cite{Vehicle}\cite{RIS}.

When the transmitter consists of a high-density LED array, VLC-based systems, particularly in traffic and vehicular environments, face significant challenges in reliable signal detection \cite{focus}. Due to focal shift, limited sensor resolution, or motion blur from camera movement, 
multiple optical signals may physically overlap on the receiver surface, creating what we refer to as 
\textit{spatial-domain inter-symbol interference (ISI)}.
This ISI makes it difficult for traditional pixel-based or threshold-based detection algorithms to distinguish between valid and interfering signals \cite{blur}.

To address the ISI issue, various approaches have been proposed. For instance, \cite{Spatialinterference} employs orthogonal preamble codes and frequency-domain separation to distinguish overlapping signals in rolling shutter-based VLC systems. 
For global shutter cameras, \cite{LEDGROUP} proposes grouping four LEDs to flash simultaneously as a means to manage ISI, while \cite{lowdensity} explores the use of low-density LED arrangements and the deliberate placement of non-emitting LEDs between active ones to reduce interference. Additionally, \cite{taiwan} explores spatial LED pattern and design coding strategies to minimize signal collisions. 
While these methods demonstrate effectiveness in mitigating ISI, they typically rely on reducing the transmission rate, such as through repetition coding, constrained LED actication.

\begin{figure*}[ht]
     \centering
     \includegraphics[width=0.95\linewidth]{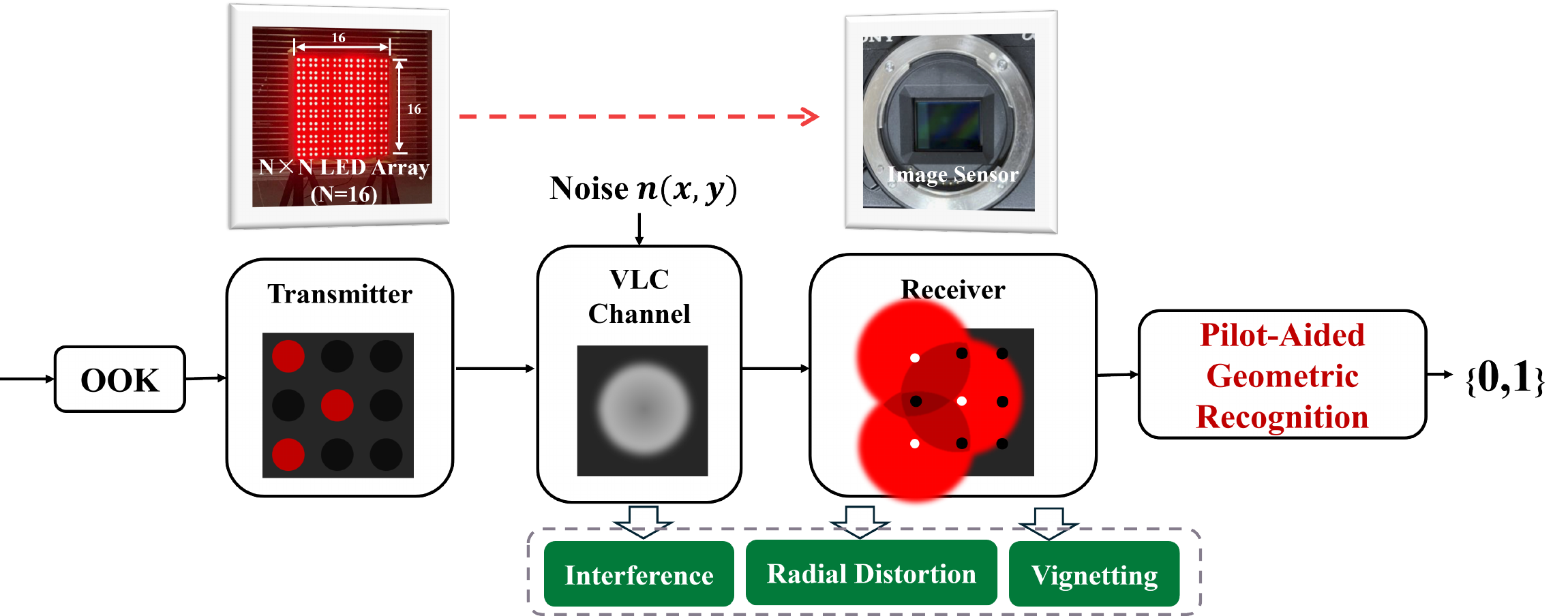}
     \caption{System model of the proposed high-density VLC system}
     \label{fig:System_model}
\end{figure*}

In addition to ISI, practical high-density VLC systems also suffer from radial distortion and vignetting effects. Radial distortion (barrel and pincushion types) geometrically deforms the projected LED grid, leading to misalignment between predicted and observed positions, which directly increases detection errors. Vignetting further complicates detection by causing edge LEDs to appear as incomplete or asymmetric spots with reduced intensity, making conventional circle detection unstable. These optical imperfections must be compensated to ensure consistent decoding performance.

In contrast, we aim to maintain full signaling density by directly extracting information from blurred and potentially overlapping LED circles. This makes robust and accurate circle detection essential for reliable decoding. 
Traditional geometric methods estimate the center of a blurred circle using chord bisectors \cite{circledectation}, but become unreliable under spatial interference due to ambiguous spot boundaries. The Hough circle transform \cite{Hough} offers better tolerance to distortion and overlap, yet often yields high \textit{false positive} rates in complex VLC scenes with multiple interfering light sources.

To address optical degradations caused by imaging systems, several studies in related fields have investigated radial distortion and lens vignetting. Radial distortion is commonly compensated using camera calibration techniques based on predefined patterns, such as chessboards or dot grids\cite{8}, to estimate distortion parameters and recover geometrically correct projections. These approaches have been widely adopted in computer vision and photogrammetry applications. However, they require explicit calibration procedures and must be repeated whenever the focal length or lens configuration changes, which limits their applicability to VLC systems operating under varying distances and optical settings.

Lens vignetting has also been extensively studied in general imaging. Model-based methods characterize vignetting as radially symmetric intensity attenuation, while data-driven approaches employ convolutional neural networks\cite{9} to learn brightness correction mappings from large datasets. Although effective in restoring image uniformity, these methods typically involve high computational complexity and are therefore unsuitable for real-time or resource-constrained VLC receivers. Consequently, existing VLC studies rarely incorporate explicit vignetting compensation and instead implicitly assume uniform illumination across the image plane.

While these optical correction techniques demonstrate effectiveness in conventional imaging tasks, they have not been systematically investigated in the context of image-sensor-based VLC. In particular, most existing VLC approaches assume an ideal pinhole camera model and neglect geometric deformation and intensity attenuation effects. Under high-density or long-distance VLC scenarios, radial distortion leads to misalignment between expected and observed LED positions, while vignetting causes edge LEDs to appear incomplete or asymmetric with reduced intensity, significantly degrading detection reliability. As a result, the lack of optical degradation awareness remains an important limitation of current VLC systems.

To overcome these limitations, we propose a novel decoding framework that accurately recovers overlapping optical signals without compromising transmission rate. 
We first formulate a mathematical model that quantifies the spatial interference. The framework also integrates radial distortion correction and vignetting-aware circle validation, ensuring geometric consistency across the field of view. Building on this model, we propose a pilot-aided detection algorithm. Geometric priors—including blur diameter and LED layout—are extracted from pilot frames and used to guide a PSF-constrained Hough transform with subsequent center alignment refinement. This enables accurate spot-level separation and significantly improves decoding robustness under severe interference.

The key contributions of this work are as follows:
\begin{itemize}
    \item We give the first VLC detection framework that directly recover information from spatially overlapped blurred spots under ISI, without reducing signal density.
    \item We introduce a pilot-aided PSF-constrained circle detection algorithm that significantly reduces false positives and improves decoding accuracy.
    \item We validate the proposed method through \textit{real-world experiments} on an image sensor-based VLC testbed, demonstrating substantial improvements in both detection accuracy and throughput compared to conventional Hough-based and low-density baseline methods.
    \item We extend the framework to address practical optical imperfections by incorporating radial distortion correction and vignetting-aware compensation, which further enhance geometric consistency and robustness at the image periphery.
\end{itemize}

\begin{table}[t]
\centering
\caption{List of Symbols}
\label{tab:notation_core}
\renewcommand{\arraystretch}{1.05}
\begin{tabular}{p{0.18\linewidth} p{0.72\linewidth}}
\toprule
\textbf{Symbol} & \textbf{Description} \\
\midrule

$N$ & Number of LEDs per dimension in the $N \times N$ LED array. \\
$(i,j)$ & Index of LED at row $i$ and column $j$. \\
$t(i,j)$ & Transmitted OOK bit from LED $(i,j)$, $t(i,j)\in\{0,1\}$. \\
$(a_i,b_j)$ & Ideal pixel coordinate of LED $(i,j)$. \\
$(a_i',b_j')$ & Distorted pixel coordinate of LED $(i,j)$ . \\
$\Delta_a,\Delta_b$ & Inter-LED pixels in horizontal and vertical directions. \\
$G$ & Full LED grid point set on the image plane. \\
$C$ & Blur diameter of LED spot (pixels). \\
$B_R(a_i,b_j)$ & Blur region (PSF support) centered at $(a_i,b_j)$ with diameter $C$. \\
$p_{(a_i,b_j)}(x,y)$ & PSF intensity contributed by LED $(i,j)$ at pixel $(x,y)$. \\

$f$ & Lens focal length. \\
$F$ & Lens F-number (aperture value). \\
$D$ & Effective aperture diameter, $D=f/F$. \\
$P$ & Pixel pitch of the image sensor. \\
$s$ & Focusing distance of the camera. \\
$s'$ & Actual communication distance (LED plane to camera). \\

$r_{i,j}$ & Radial distance of LED $(i,j)$ from the image center. \\
$r_{\max}^{\text{pixel}}$ & Maximum usable image radius without vignetting. \\
$\eta_{i,j}$ & Visible-area ratio under vignetting ($0\le\eta_{i,j}\le1$). \\

$\mathbb{K}=\{k_1,k_2\}$ & Radial distortion parameters. \\

$r(x,y)$ & Received pixel intensity (superposition of PSFs + noise). \\
$I_{(a_i,b_j)}$ & Average intensity within region $B_R(a_i,b_j)$ for detection. \\
$\gamma$ & Detection threshold for OOK symbol recovery. \\

$\epsilon$ & Radius tolerance factor in PSF-constrained Hough detection. \\
$\theta$ & Threshold for matching detected centers to grid points. \\

\bottomrule
\end{tabular}
\end{table}

\section{System model}

\subsection{Transmitter}

We consider a VLC system that consists of a high-density LED array transmitter, a VLC channel, and an image sensor receiver illustrated in \figurename~\ref{fig:System_model}. The LEDs lie on a $N \times N$ array plane uniformly spaced. Let $(i,j)$ be the spatial coordinates of LEDs with  $i,j \in\{0, 1, \dots, N-1\}$. Each LED emits modulated light signals $t(i,j) \in \{0,1\}$ independently using On-Off Keying (OOK).

\subsection{Blurred LED spot model (PSF support)}
Assuming that $(a_i, b_j) \in G$ is the coordinate of captured LED $(i,j)$ on the image sensor, where the LED grid point set $G$ is defined as 
\begin{equation}
\label{eq:G}
 G = \big\{(a_i, b_j)\;\big|\; a_i = a_{BL} + i \cdot \Delta_a,~ b_j = b_{BL} + j \cdot \Delta_b\big\}
\end{equation}
where $({a}_{BL} ,~ {b}_{BL})$ is the bottom-left LED coordinate and $\Delta_a,\Delta_b$ are the pixel distance between adjacent LEDs on the captured image, determined by the actual object distance, lens focal length, and sensor resolution.

Due to defocus and lens/sensor limitations, each LED is captured as a blurred circular spot. 
We adopt a  point spread function (PSF)-support model: 
\begin{equation}
p_{(a_i,b_j)}(x,y) =
\begin{cases}
h \cdot t(i,j), & \text{if } (x, y) \in B_R(a_i,b_j) \\
0, & \text{otherwise}
\end{cases}
\end{equation}
where $h$ is the energy attenuation function normalized to $[0,1]$ that decreases with distance from the center and $B_R(a_i,b_j)$ is a circular region centered at $(a_i,b_j)$ with blur diameter $C$ as  
\begin{equation}
B_R(a_i,b_j) = \{ (x, y) \mid (x - a_i)^2 + (y - b_j)^2 \leq (\frac{C}{2})^2 \}
\end{equation}
Note that the blur diameter $C$ is not constant but fluctuates depending on the optical and sensor-related factors.

\subsection{Image formation and baseline detection}
Assuming perfect  synchronization between transmitter and receiver, the total received signal at sensor coordinate $(x, y)$ is modeled as the superposition of all active LED contributions.

Let \( \mathcal{A} = \{(i,j) \mid t(i,j) = 1\} \) be the set of active LEDs. Due to the blur diameter \( C \) exceeding the inter-LED spacing \( \Delta_a, \Delta_b \), multiple PSFs can simultaneously affect a pixel. The received signal is

\begin{eqnarray}
\label{eq:r}
r(x,y) = \sum_{a_i} \sum_{b_j} p_{({a_i},{b_j})}(x,y) + n(x,y) \nonumber \\
=\sum_{(i,j) \in \mathcal{A}} p_{(a_i,b_j)}(x,y)
+ n(x,y)
\end{eqnarray}
where $n(x,y)$ denotes additive ambient noise. The image sensor samples $r(x,y)$ to form a 2D intensity matrix $\mathbf{R}$.

The objective of the receiver is to recover the transmitted binary matrix $\mathbf{T} = \{t(i,j)\}$ from the captured image $\mathbf{R}$, even in the presence of spot blur, spatial overlap, and background noise.

To decode each transmitted bit $t(i,j)$ from the received image $\mathbf{R}$, the receiver first estimates the blur diameter $C$ based on system calibration or online measurements, which defines the circular support region $B_R(a_i, b_j)$ for each LED $(i,j)$. The detection statistic $I_{(a_i,b_j)}$ is computed as the average pixel
intensity within this region:
\begin{equation}
I_{(a_i,b_j)} = \mathrm{mean}\,\{\, r(x,y) \mid (x,y)\in B_R(a_i,b_j) \,\}.
\end{equation}

A fixed threshold $\gamma$ is applied to determine the transmitted bit:
\begin{equation}
\hat{t}(i,j) =
\begin{cases}
1, & I_{(a_i,b_j)} > \gamma \\
0, & \text{otherwise}
\end{cases}
\end{equation}
As a result, the transmitted binary matrix $\hat{\mathbf{T}} = \{\hat{t}(i,j)\}$ is decoded from the received signal.

\subsection{Impairments: radial distortion, spatial ISI, and vignetting}

\begin{figure}[!t]
     \centering
     \includegraphics[width=0.95\linewidth]{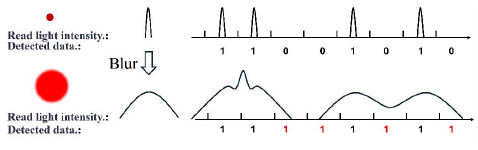}
     \caption{Inter-symbol interference}
     \label{fig:Inter-symbol interference}
\end{figure}

\textbf{Spatial-domain ISI} arises when multiple active LEDs overlap in high-density LED arrays, where spacing between neighboring LEDs is small. \figurename~\ref{fig:Inter-symbol interference} shows a typical example of one-dimension ISI. 


\begin{figure}[htbp]
\centering
\begin{tabular}{cc}
    (a)\ \includegraphics[width=0.35\linewidth]{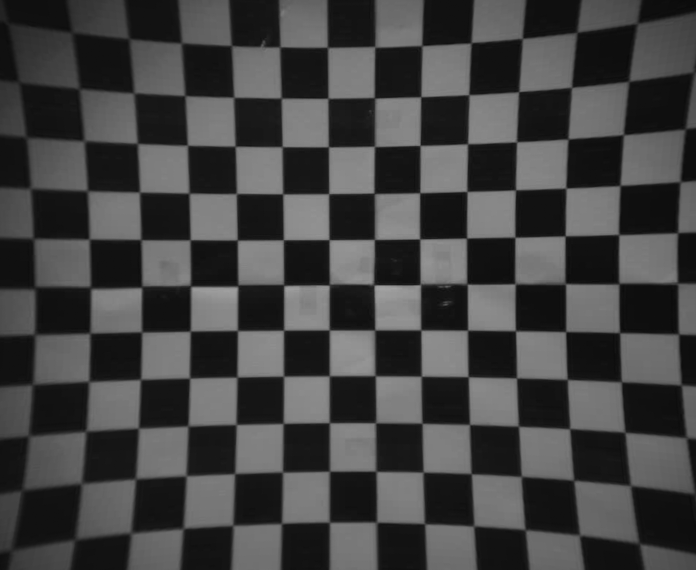} &
    (b)\ \includegraphics[width=0.35\linewidth]{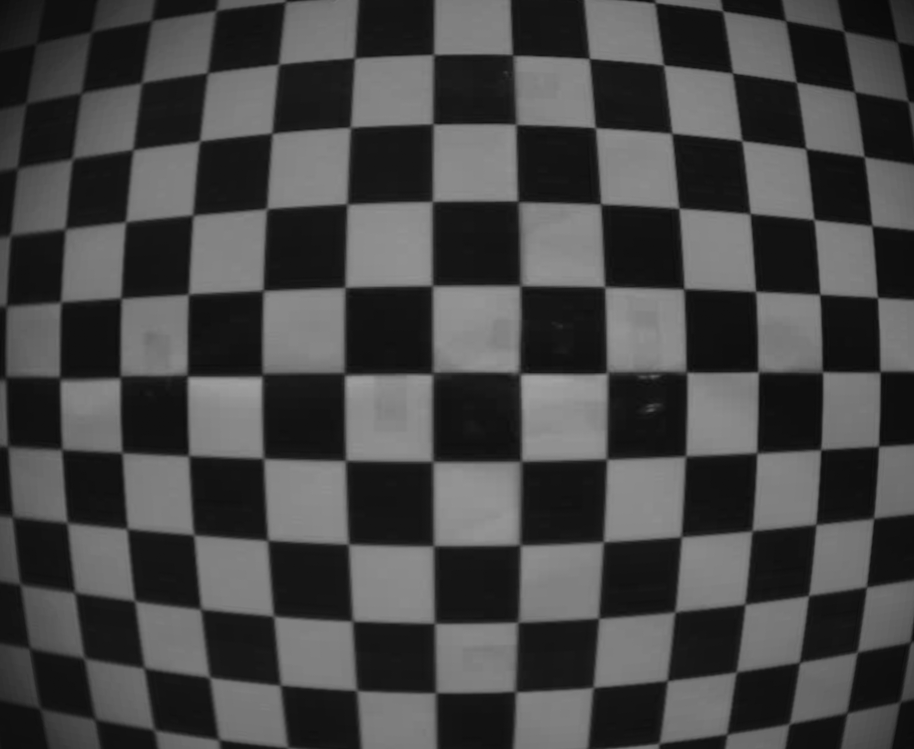} \\
    (c)\ \includegraphics[width=0.35\linewidth]{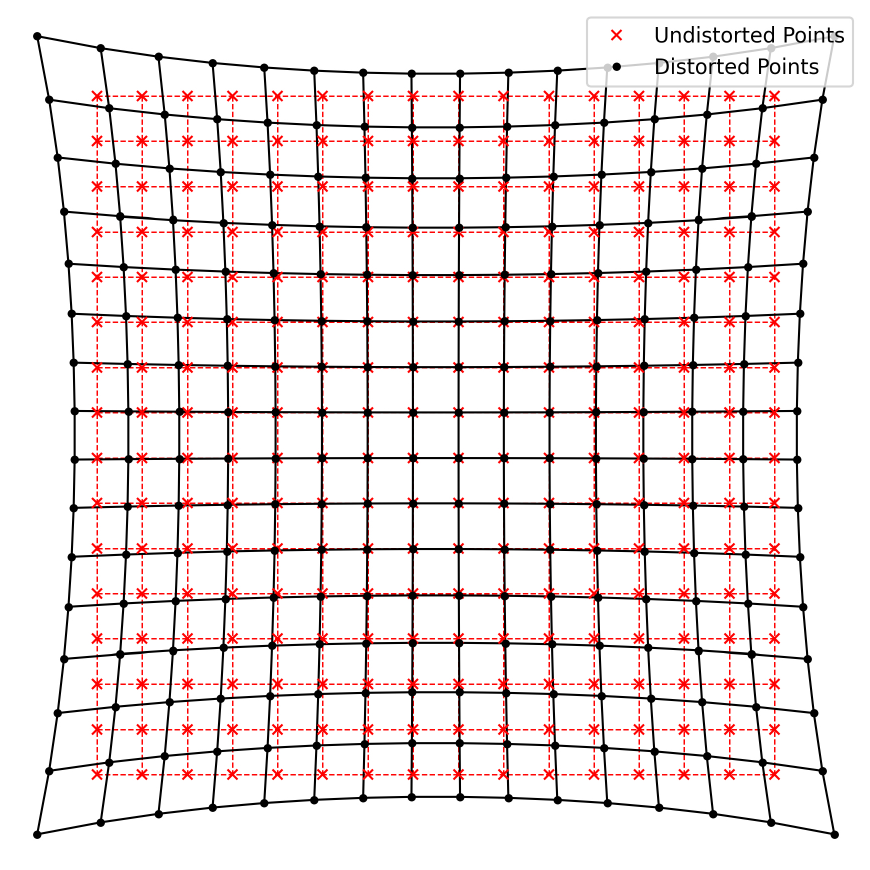} &
    (d)\ \includegraphics[width=0.35\linewidth]{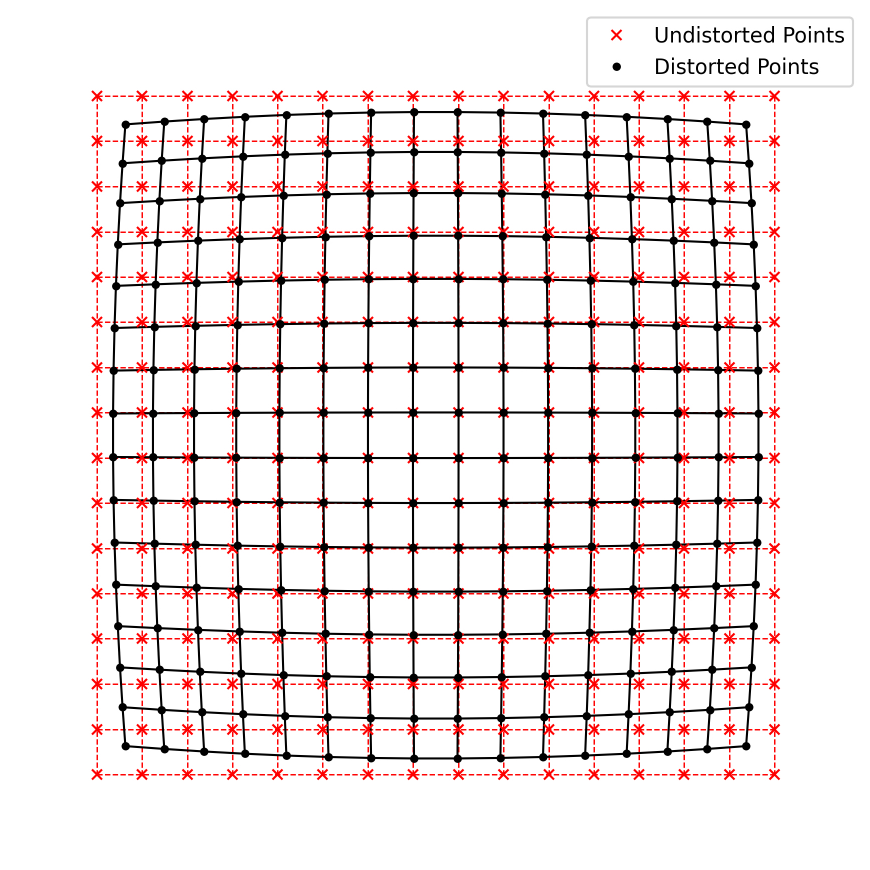}
\end{tabular}
\caption{Radial Distortion:(a)(c) Pincushion Distortion; (b)(d) Barrel Distortion.}
    \label{fig:Distortion}
\end{figure}

\textbf{Radial distortion} as shown in \figurename~\ref{fig:Distortion}, shifts the apparent spot centers to $(a'_i,b'_j)$, the corresponding received signal becomes 
\begin{equation}
r_{\mathrm{RD}}(x,y) =
\sum_{a_i'} \sum_{b_j'} p_{(a_i', b_j')}(x, y)
+ n(x, y)
\end{equation}


In particular, the Hough-detected center may fall outside the detection threshold region of the predicted grid point, causing missed detections or false positives. Therefore, a robust distortion correction step is essential to re-align the captured LED layout with its ideal configuration, ensuring consistency between the signal reception location and the expected LED grid coordinates.



\begin{figure}[t!]
\centering
(a)\includegraphics[width=0.32\linewidth]{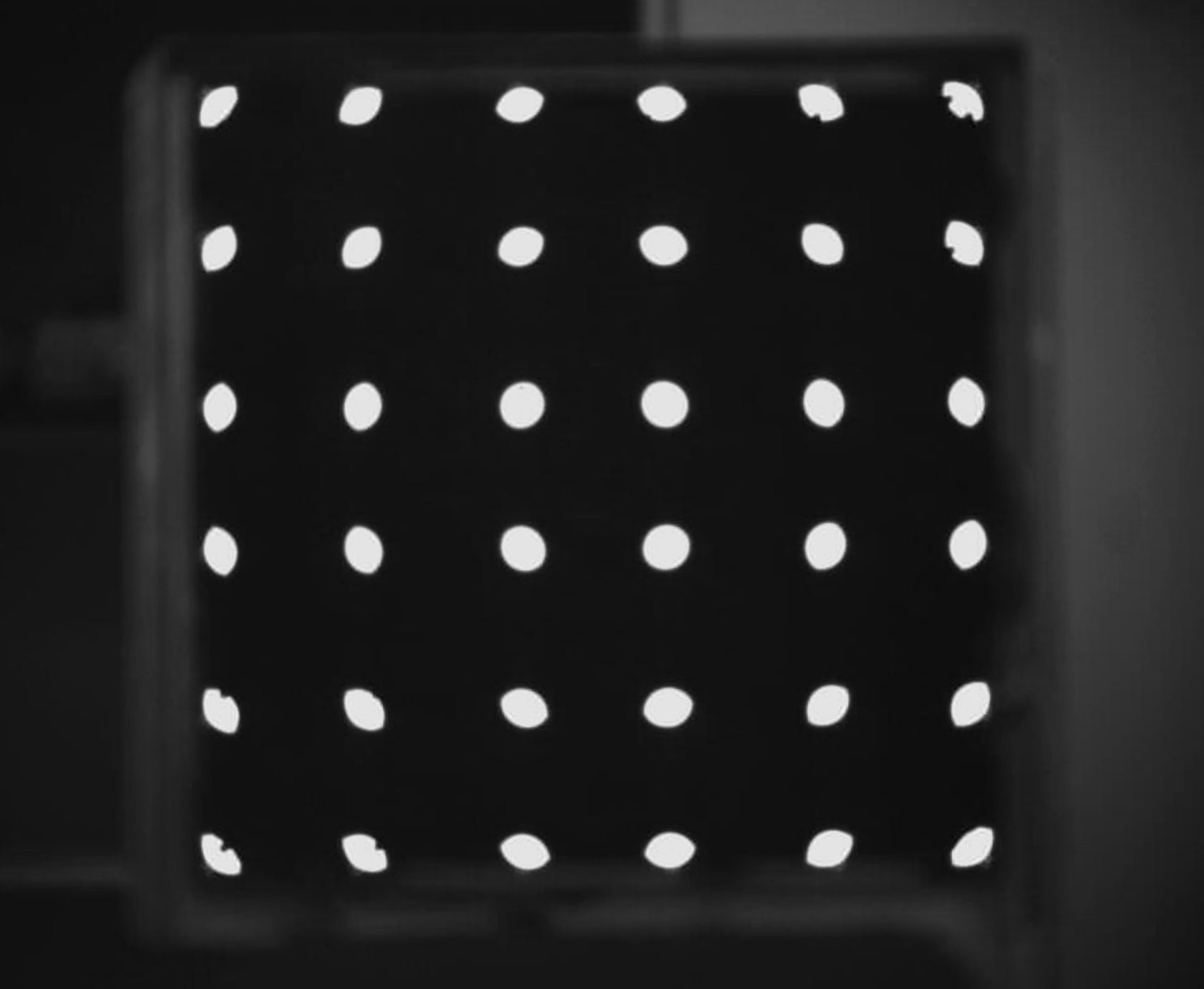}
\hfill
(b)\includegraphics[width=0.5\linewidth]{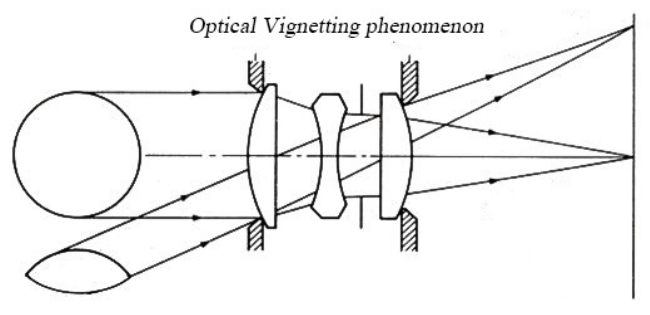}
\caption{Lens Vignetting Phenomenon:(a) Experiment Figure; (b) Schematic Diagram.}
\label{fig:Vignetting}
\end{figure}

\subsection{Radial Distortion}

For the Radial Distortion, a robust distortion correction step is essential to re-align the captured LED layout with its ideal configuration, ensuring consistency between the signal reception location and the expected LED grid coordinates.

The Received distorted PSF is then centered at $(a_i', b_j')$ with support:
\begin{equation}
B_R(a_i', b_j') = \{(x, y) \mid (x - a_i')^2 + (y - b_j')^2 \le (\tfrac{C}{2})^2\},
\end{equation}
\begin{equation}
p_{(a_i', b_j')}(x, y) =
\begin{cases}
h \cdot t(i,j), & (x, y) \in B_R(a_i', b_j'),\\[3pt]
0, & \text{otherwise},
\end{cases}
\end{equation}

its distorted projection $(a_i',b_j')$ using the radial distortion model:
\begin{align}
(a_i', b_j') 
=& f_{\text{distort}}(a_i, b_j; \mathbb{K})\\
= \bigl(
    &c_x + (a_i - c_x)\bigl(1 + k_1 r_{i,j}^2 + k_2 r_{i,j}^4 + k_3r_{i,j}^6+~...\bigr),
    \nonumber\\
    &c_y + (b_j - c_y)\bigl(1 + k_1 r_{i,j}^2 + k_2 r_{i,j}^4 + k_3r_{i,j}^6+~...\bigr)
\label{eq:radial_distort_no_xy}
\end{align}

where $\mathbb{K}=\{k_1,k_2,k_3,~...\}$ denotes the radial distortion parameter. It characterizes the intrinsic radial distortion of the imaging system which is:
\begin{itemize}
  \item \textbf{Independent of:}
  \begin{itemize}
    \item the communication distance;
    \item pixel indices $(i,j)$;
    \item the identity or indexing of individual LEDs or feature points;
    \item exposure time and frame rate.
  \end{itemize}

  \item \textbf{Dependent on:}
  \begin{itemize}
    \item focal length;
    \item lens replacement or lens type;
    \item changes in the effective optical configuration of the camera.
  \end{itemize}
\end{itemize}
For a given camera with a fixed optical configuration, these parameters can be regarded as global constants shared across the entire image plane. 
The sign and magnitude of the first-order coefficient $k_1$ provide intuitive insight into the distortion type: $k_1 < 0$ typically corresponds to barrel distortion, whereas  $k_1 > 0$ indicates pincushion distortion. 
Higher-order parameters are mainly introduced to compensate for residual nonlinear distortion effects in peripheral image regions.

$(c_x, c_y)$ is the camera principal point  in the distorted imageand and $r_{i,j}$ is the normalized radial distance written as

\begin{equation}
    r_{i,j}^2
    =
    \left(\frac{a_i - c_x}{f}\right)^2
    +
    \left(\frac{b_j - c_y}{f}\right)^2 .
    \label{eq:r_ij_normalized_no_xy}
\end{equation}

Finally, the total received signal affected by radial distortion is expressed as:
\begin{equation}
\boxed{
r_{\mathrm{RD}}(x,y) =
\sum_{a_i'} \sum_{b_j'} p_{(a_i', b_j')}(x, y)
+ n(x, y)
}
\end{equation}

\subsection{Vignetting Challenges}

As is shown in the \figurename~\ref{fig:Vignetting}, in wide-field lenses or light-field camera, light spots or sub-images near the image periphery deviate from perfect circular shapes and often appear as crescent or ``cat's-eye'' patterns. This phenomenon arises because light emitted from edge LEDs enters the lens at large incident angles; only part of the light cone passes through the finite lens aperture while the rest is blocked.

The maximum unvignetted chief ray angle is determined by\cite{vignetting_correction1}
\begin{equation}
\theta_{\max} = \arctan \left( \frac{D}{2L} \right), \qquad D = \frac{f}{F}
\label{eq:chief_ray_angle}
\end{equation}
where \(D\) denotes the effective aperture, \(f\) is the focal length, \(F\) is the F-number, and \(L\) is the lens length. The corresponding maximum effective image radius is
\begin{equation}
r_{\max} = f \tan(\theta_{\max}), \qquad r_{\max}^{\text{pixel}} = \frac{r_{\max}}{P}
\label{eq:r_max}
\end{equation}
where \(r_{\max}^{\text{pixel}}\) is the radius in image coordinates, and \(P\) is the pixel size.

When the radial distance \(r_{i,j}\) of LED $(i,j)$ exceeds \(r_{\max}^{\text{pixel}}\), its spot is partially clipped. The radial exceedance is computed as
\begin{equation}
\Delta_{i,j} = r_{i,j} - r_{\max}^{\text{pixel}}.
\end{equation}

The visible area of the spot under partial occlusion is modeled geometrically. The visible crescent area for a circular aperture clipped by \(\Delta_{i,j}\) is given by:
\begin{equation}
A(\Delta_{i,j})
= \left(\frac{C}{2}\right)^2
\!\arccos\!\Bigl(\frac{2\Delta_{i,j}}{C}\Bigr)
- \Delta \sqrt{\left(\frac{C}{2}\right)^2 - \Delta_{i,j}^2},
\end{equation}
\begin{equation}
\Gamma(\Delta_{i,j})=
\frac{A(\Delta_{i,j})}{\pi (C/2)^2},
\end{equation}
which defines the fractional visible area compared to the full PSF.

Thus, the normalized visible-area ratio $\eta_{i,j}$ for each LED spot becomes:
\begin{equation}
\eta_{i,j}=
\begin{cases}
1, & \Delta_{i,j} \le 0,\\[3pt]
\Gamma(\Delta_{i,j}), & \Delta_{i,j} > 0.
\end{cases}
\end{equation}

We refine the PSF model by applying $\eta_{i,j}$ and a spatially-varying attenuation $h_{\eta_{i,j}}$ that captures optical degradation effects at oblique angles:
\begin{equation}
p^{(\mathrm{V})}_{(a_i,b_j)}(x,y)=
\begin{cases}
\eta_{i,j} \cdot h_{\eta_{i,j}}\cdot t(i,j), & (x,y)\in B_R(a_i,b_j),\\[3pt]
0, & \text{otherwise}.
\end{cases}
\end{equation}

Consequently, the vignetting-aware received signal becomes:
\begin{equation}
r_{\mathrm{V}}(x,y)=
\sum_{a_i}\sum_{b_j} p^{(\mathrm{V})}_{(a_i,b_j)}(x,y)
+ n(x,y)
\end{equation}

Here, $\eta_{i,j} \in [0,1]$ reflects the degree of visible PSF clipping due to vignetting. This formulation captures the spatially varying brightness loss in peripheral LEDs and provides the analytical basis for compensation and robust symbol detection under optical constraints.

This formulation models the spatially varying attenuation of LED blur intensity caused by vignetting and establishes the analytical foundation for brightness compensation strategies presented in Section~\ref{sec:Vignetting Correction}.

\subsection{Unified Received Signal Model}

To jointly capture the effects of radial distortion, spatial-domain interference, and vignetting attenuation in a unified framework, we define the final received signal model as:
\begin{equation}
\label{eq:r_all}
r_{\mathrm{ALL}}(x,y)=
\sum_{a_i'}\sum_{b_j'}
p^{(\mathrm{ALL})}_{(a_i',b_j')}(x,y)
+ n(x,y)
\end{equation}
where
\begin{equation}
p^{(\mathrm{ALL})}_{(a_i',b_j')}(x,y)=
\begin{cases}
\eta_{i,j} \cdot h \cdot t(i,j),
& (x,y)\in B_R(a_i',b_j'),\\[3pt]
0, & \text{otherwise}.
\end{cases}
\end{equation}
which integrates (i) geometric displacement from radial distortion, (ii) PSF overlap from ISI, and (iii) attenuation/clipping from vignetting.

\section{Pilot Phase: Grid and Blur Estimation via PSF-Aided Calibration}

\begin{figure}[!h]
\centering
\includegraphics[width=\linewidth]{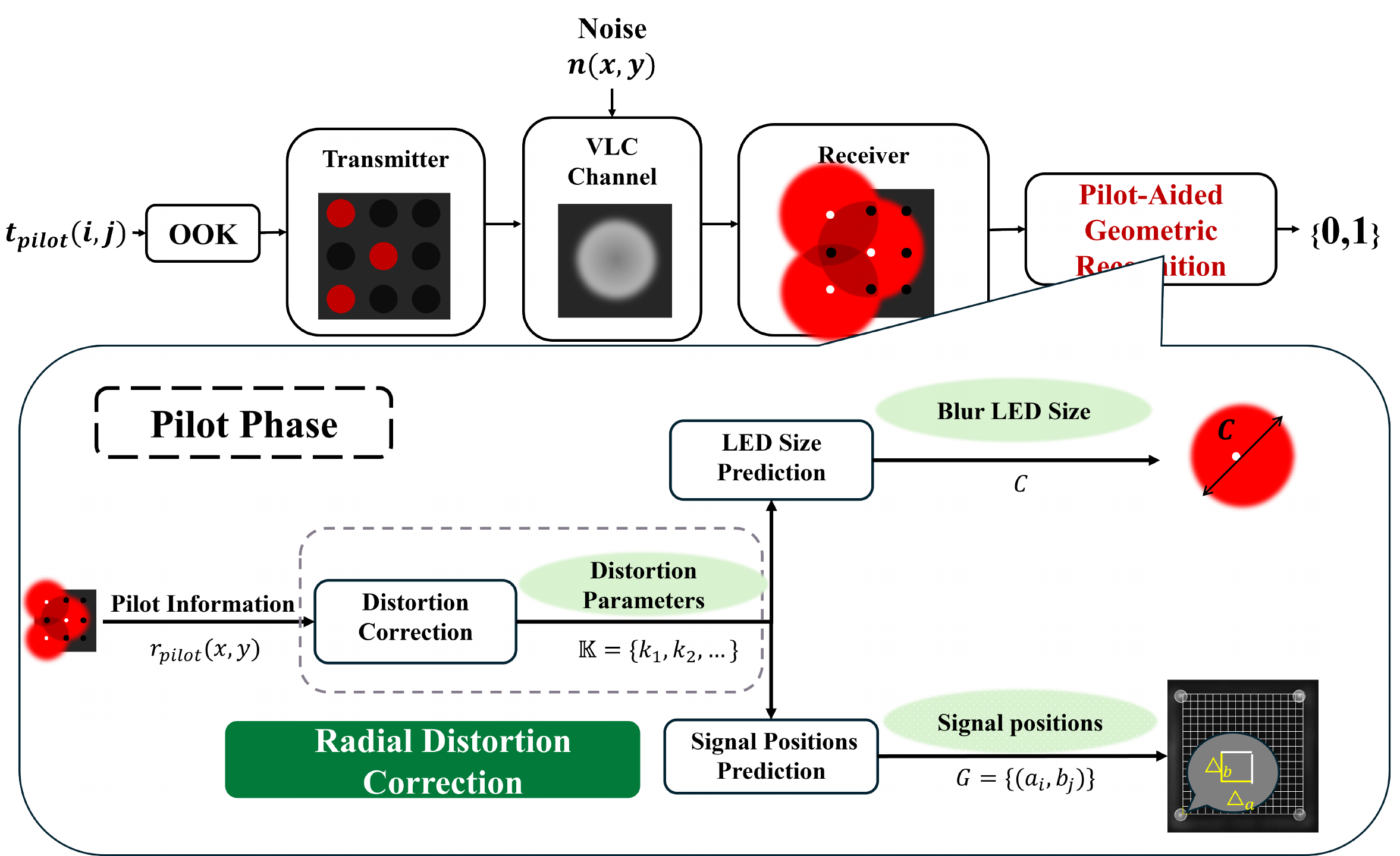} 
\caption{The Method for Detecting Blur LEDs---Pilot Phase}
\label{fig:method1}
\end{figure}

To address inter-symbol interference (ISI) caused by overlapping blurred LED spots in high-density visible light communication (VLC) systems, we propose a robust two-stage decoding framework. The first stage, referred to as the pilot phase, focuses on geometric calibration by correcting lens-induced distortions. As illustrated in \figurename~\ref{fig:method1}, this phase estimates the LED grid layout and blur size using pilot signals. Specifically, the pilot frame is employed to recover the ideal LED coordinates, determine the blur diameter, and locate the signal positions that serve as geometric priors for subsequent decoding.

\subsection{Pilot information Definition}
High-density LED array image sensor communication systems require precise grid alignment for accurate decoding. Lens distortion and geometric misalignment between the transmitter and camera introduce significant errors in the detected signal positions, such as radial distortion that causes straight lines to appear curved, tangential distortion that shifts points asymmetrically, and perspective distortions from transmitter tilt that warp the regular grid into a trapezoidal shape. These effects degrade decoding accuracy by misaligning the LED grid structure in the captured image. 

Let $t_{\mathrm{pilot}}(i,j)\in\{0,1\}$ denote the known pilot activation pattern
on the $N\times N$ LED array. The received pilot frame on the image sensor is
modeled as
\begin{equation}
    r_{\mathrm{pilot}}(x,y)
    =
    \sum_{(a'_i,b'_j)}
    p^{(\mathrm{pilot})}_{(a'_i,b'_j)}(x,y)
    + n(x,y),
    \label{eq:r_pilot_def}
\end{equation}
where 
$p^{(\mathrm{pilot})}_{(a'_i,b'_j)}(x,y)$ is the pilot-induced PSF contribution
from LED $(i,j)$, given by
\begin{equation}
    p^{(\mathrm{pilot})}_{(a'_i,b'_j)}(x,y)
    =
    \begin{cases}
        \eta_{i,j}\,h\,t_{\mathrm{pilot}}(i,j), 
        & (x,y)\in BR(a'_i,b'_j),\\[2mm]
        0, & \text{otherwise},
    \end{cases}
    \label{eq:p_pilot_psf}
\end{equation}
with $(a'_i,b'_j)$ denoting the radially distorted LED centers, 
$\eta_{i,j}\in[0,1]$ the vignetting-induced visible-area ratio,
$h$ the normalized PSF amplitude.

A sparse pilot LED pattern is chosen such that the distance between any two pilot signals satisfies
\begin{equation}
\|(a_i,b_j) - (a_{i'},b_{j'})\| > C,
\qquad \forall (i,j)\neq(i',j').
\label{eq:pilotDistance}
\end{equation}
The corresponding ideal undistorted pilot grid point set is given by
\begin{equation}
G_{\text{pilot}}
= \{\, (a_i, b_j) \mid (i,j)\text{ satisfy eq } \eqref{eq:pilotDistance} \,\}.
\end{equation}

In the considered 
 $16 \times 16$ LED array configuration, the pilot LEDs are uniformly selected at fixed intervals along both horizontal and vertical directions. Specifically, the pilot grid is given by
\begin{equation} 
G_{\text{pilot}} = \left\{ (a_i, b_j) \mid i,j \in \{0,3,6,9,12,15\} \right\}, 
\label{eq:G_pilot}
\end{equation}

\subsection{Radial Distortion Correction}
\subsubsection{Estimation of distortion coefficient}

Among all detected pilot LEDs, we automatically select as the reference pilot the one closest to the camera principal point $(c_x,c_y)$ in the distorted image. Let $G_{\text{pilot}}$ denote the index set of pilot LEDs and $(a'_{i,j}, b'_{i,j})$ their distorted coordinates. The reference pilot $(i_0,j_0)$ is chosen as
\begin{equation}
(i_0,j_0)
=
\arg\min_{(i,j)\in G_{\text{pilot}}}
\bigl[(a'_{i,j}-c_x)^2 + (b'_{i,j}-c_y)^2\bigr].
\label{eq:ref_pilot_selection_final}
\end{equation}
This distorted reference coordinate undergoes radial-undistortion and homography-based rectification, yielding the ideal reference position $(a_{\mathrm{ref}}, b_{\mathrm{ref}})$. The residual distortion after
rectification is negligible, and $(a_{\mathrm{ref}}, b_{\mathrm{ref}})$ is used as the grid origin in the ideal projection.

Since the LED array has uniform physical spacing of $3\,\mathrm{cm}$, the projected spacing between adjacent LEDs on the ideal image plane is
\[
\Delta_a=\Delta_b=\alpha,
\]
where the geometric scaling factor $\alpha$ is given by
\begin{equation}
    \alpha = \frac{f}{p} \cdot \frac{3\,\mathrm{cm}}{s'},
\end{equation}
with $f$ the focal length, $p$ the pixel pitch, and $s'$ the distance between
the LED plane and the camera.

Consequently, the ideal undistorted grid coordinates are parameterized as
\begin{equation}
    a_i = a_{\mathrm{ref}} + \alpha (i - i_0),
    \qquad
    b_j = b_{\mathrm{ref}} + \alpha (j - j_0),
    \label{eq:ideal_grid_ref}
\end{equation}

\label{sec:Lens Distortion Correction}

To model Radial distortion, we express the radial distance directly in terms of pixel coordinates and focal length. Let assume $f_x \approx f_y = f$. The normalized radial distance is then written as
\begin{equation}
    r_{i,j}^2
    =
    \left(\frac{a_i - c_x}{f}\right)^2
    +
    \left(\frac{b_j - c_y}{f}\right)^2 .
    \label{eq:r_ij}
\end{equation}
Each ideal LED coordinate $(a_i,b_j)$ is mapped to its distorted projection $(\hat{a_i'},\hat{b_j'})$ using the radial distortion model:

\begin{equation}
\begin{aligned}
(\hat{a_i'}, \hat{b_j'}) 
=& f_{\text{distort}}(a_i, b_j; \mathbb{K})\\
= &\bigl(c_x + (a_i - c_x)\bigl(1 + k_1 r_{i,j}^2 + k_2 r_{i,j}^4+~...\bigr)
\\
    &\bigl(c_y + (b_j - c_y)\bigl(1 + k_1 r_{i,j}^2 + k_2 r_{i,j}^4+~...\bigr)
\label{eq:radial_distort}
\end{aligned}
\end{equation}

where $\mathbb{K}=\{k_1,k_2, \cdots \}$ denotes the radial distortion parameters.

The optimal distortion parameters are obtained by minimizing the total squared error between observed$(a_i',b_j')$ and predicted distorted$(\hat{a_i'}, \hat{b_j'})$ coordinates:
\begin{equation}
\min_{\mathbb{K}} \sum_{(i,j)} \left\|
\begin{pmatrix}
a_i' \\
b_j'
\end{pmatrix}
-
\begin{pmatrix}
\hat{a_i'}\\
\hat{b_j'}
\end{pmatrix}
\right\|^2.
\label{eq.a-b}
\end{equation}

According to eq\eqref{eq:r_ij_normalized_no_xy}, the radial distance is defined in normalized camera coordinates by scaling the pixel displacement with the focal length $f$. 
As a result, the radial distortion parameters$\mathbb{K}=\{k_1,k_2, \cdots \}$are dimensionless and their magnitudes are independent of the image resolution and pixel pitch. 
Under this normalization, the radial coordinate typically satisfies $r \in [0,1]$ for most image regions, and slightly exceeds unity near the image corners for wide field-of-view cameras.  
Larger values may be observed when strong edge distortion is present or when higher-order distortion terms are truncated.

These distortion parameters will be used in the Information Phase(Section~\ref{sec:Ideal_Coordinates}).

\subsubsection{Estimation of the ideal coordinates based on $\mathbb{K}$ for pilot}

In the pilot phase, we first use $\mathbb{K}=\{k_1,k_2, \cdots \}$ to estimate the ideal coordinates which are used for signal position estimation and blur LED size estimation.

After estimating the distortion parameters \( \mathbb{K} = \{k_1, k_2\} \), the corrected LED grid positions are obtained by applying the inverse distortion model, yielding the distortion-compensated coordinates $(\hat{a_i}, \hat{b_j})$. 

The inverse radial-distortion approximation directly maps the distorted pixel coordinates $(a_i', b_j')$ to the rectified coordinates $(\hat a_i, \hat b_j)$ as
\begin{equation}
r_d^2 = 
\left(\frac{a_i' - c_x}{f}\right)^2 +
\left(\frac{b_j' - c_y}{f}\right)^2 
\label{eq.r_d}
\end{equation}

\begin{equation}
\begin{aligned}
\hat a_i = 
c_x + 
\frac{a_i' - c_x}{1 + k_1 r_d^2 + k_2 r_d^4+~...}
\\
\hat b_j = 
c_y + 
\frac{b_j' - c_y}{1 + k_1 r_d^2 + k_2 r_d^4+~...}
\end{aligned}
\end{equation}

This expression corresponds to the first-order inverse radial model, where the undistorted radius is approximated by the distorted radius $r_d$.

For notational simplicity, the rectified coordinates are henceforth denoted as
$(a_i,b_j)$, i.e., we make the approximation
\begin{equation}
(\hat{a}_i,\hat{b}_j) \approx (a_i,b_j),
\end{equation}
since the residual distortion after radial correction is negligible.

\subsection{Signal Positions Estimation}
\subsubsection{Estimation of LED grid point set $G$}
According to eq\eqref{eq:G_pilot}, the geometric center of the LED Array is 
\begin{equation}
\label{eq:center}
(a_C, b_C) = \frac{1}{36} \sum_{(a_i, b_i) \in G_{pilot} } (a_i, b_j)
\end{equation}

The four farthest points from the center \((a_C, b_C)\) within the set \(G_{\text{pilot}}\) are selected as:

\begin{equation}
\label{eq:farthest_four}
\{(a^*_k, b^*_k)\}_{k=1}^4 = \underset{(a_i,b_j) \in G_{\text{pilot}}}{\operatorname{arg\,top\,4}} \; (a_i - a_C)^2 + (b_j - b_C)^2.
\end{equation}

Each selected farthest point \( (a^*, b^*) \) is then assigned to one of the four corner roles(top-left, top-right, bottom-left, bottom-right) based on its relative position with respect to the center:
\begin{equation}
\label{eq:corner}
(a^*,b^*) = 
\begin{cases}
(a_{TL},b_{TL}) & \text{if } a^* < a_C \text{ and } b^* > b_C, \\[6pt]
(a_{TR},b_{TR}) & \text{if } a^* > a_C \text{ and } b^* > b_C, \\[6pt]
(a_{BL},b_{BL}) & \text{if } a^* < a_C \text{ and } b^* < b_C, \\[6pt]
(a_{BR},b_{BR}) & \text{if } a^* > a_C \text{ and } b^* < b_C.
\end{cases}
\end{equation}

The grid width $W$ and height $H$ are computed as the Euclidean distances between top-left and top-right, and top-left and bottom-left points, respectively:
\begin{equation}
\label{eq:Width}
\begin{aligned}
W &= \sqrt{(a_{TR} - a_{TL})^2 + (b_{TR} - b_{TL})^2} \\
H &= \sqrt{(a_{BL} - a_{TL})^2 + (b_{BL} - b_{TL})^2}
\end{aligned}
\end{equation}

For an $N \times N$ LED array, the average inter-LED distance in image pixels is defined as:
\begin{equation}
\Delta_a = \frac{W}{N - 1}, ~\Delta_b = \frac{H}{N - 1}
\label{eq:dactual}
\end{equation}

Finally, we obtain the grid point set $G = \{(a_i, b_j)\}$ based on the $\Delta_a$ and $\Delta_b$.

\subsection{Blur LED Size Estimation}
\subsubsection{Estimation of Blur LED Size $C_{exp}$}
Assume inter-LED pixel distance $\Delta _{a_{6\text{m}}}$ is known when communication distance is 6 meters, the communication distance $s'$ can be inferred from the ratio of the reference spacing to the observed inter-LED spacing $\Delta_a$, assuming a proportional relationship:
\begin{equation}
s' = \frac{\Delta_{a_{6\text{m}}}}{\Delta_a} \cdot 6\ \text{m}
\label{eq:s'}
\end{equation}

The effective aperture \( D \) of a lens is related to its focal length \( f \) and F-number \( F \) by the following equation:
\begin{equation}
D = \frac{f}{F}
\end{equation}

The diameter of the blur LED \(  C_{\text{calc}} \)\cite{optics} can be expressed by the following formula:
\begin{equation}
\label{eq:calc}
 C_{\text{calc}}= \frac{D\cdot f \cdot |s - s'|}{s' \cdot (s - f)}
\end{equation}
where $s$ is focusing distance, $F$ is aperture value, $f$ is focal length.

To convert the theoretical value $C_{\text{calc}}$ from physical units (µm) into pixel units, the pixel pitch $P$ of the image sensor must be defined. It is given by the ratio of the physical width of the camera sensor to the pixel resolution of the captured image. Therefore, the experienced blur diameter $C_{\text{exp}}$ is estimated as
\begin{equation}
     C_{\text{exp}} = k \cdot \frac{C_{\text{calc}}}{P}
\label{eq:C}
\end{equation}
with correction factor $k \approx 1.3$ in this paper.

\textbf{Remark:}
This distortion correction framework compensates not only for lens-induced nonlinear geometric deformation but also for the tilt of the transmitter plane. By integrating pilot-signal-based calibration with homography estimation, the system maintains consistent grid alignment and achieves low error rates even under practical deployment conditions involving arbitrary transmitter orientations. Moreover, this calibration enhances the robustness of signal interpretation across different lenses and focal lengths, establishing a 
reliable geometric foundation for subsequent signal demodulation.

Once the distortion-corrected grid point set $G$ is generated from the pilot image, conventional ISC systems typically determine binary symbols by sampling the light intensity at each grid location. However, in high-density LED arrays, substantial optical blur causes the light emitted from a single LED to spread over a wide region, introducing interference at adjacent grid points. As a result, signal values can no longer be reliably inferred solely from the predefined grid set $G$, and conventional sampling-based demodulation becomes highly error-prone.

To overcome this limitation, we propose a robust circle-detection-based signal extraction method specifically designed for high-density VLC scenarios with severe inter-symbol interference (ISI). By leveraging PSF-constrained circle detection and center verification, the proposed method accurately locates the dominant blur centers corresponding to each active LED and enables reliable symbol recovery even when adjacent PSF regions exhibit significant overlap.

\section{Information Phase: Vignetting and Interference Correction}
\begin{figure}[!h]
\centering
\includegraphics[width=\linewidth]{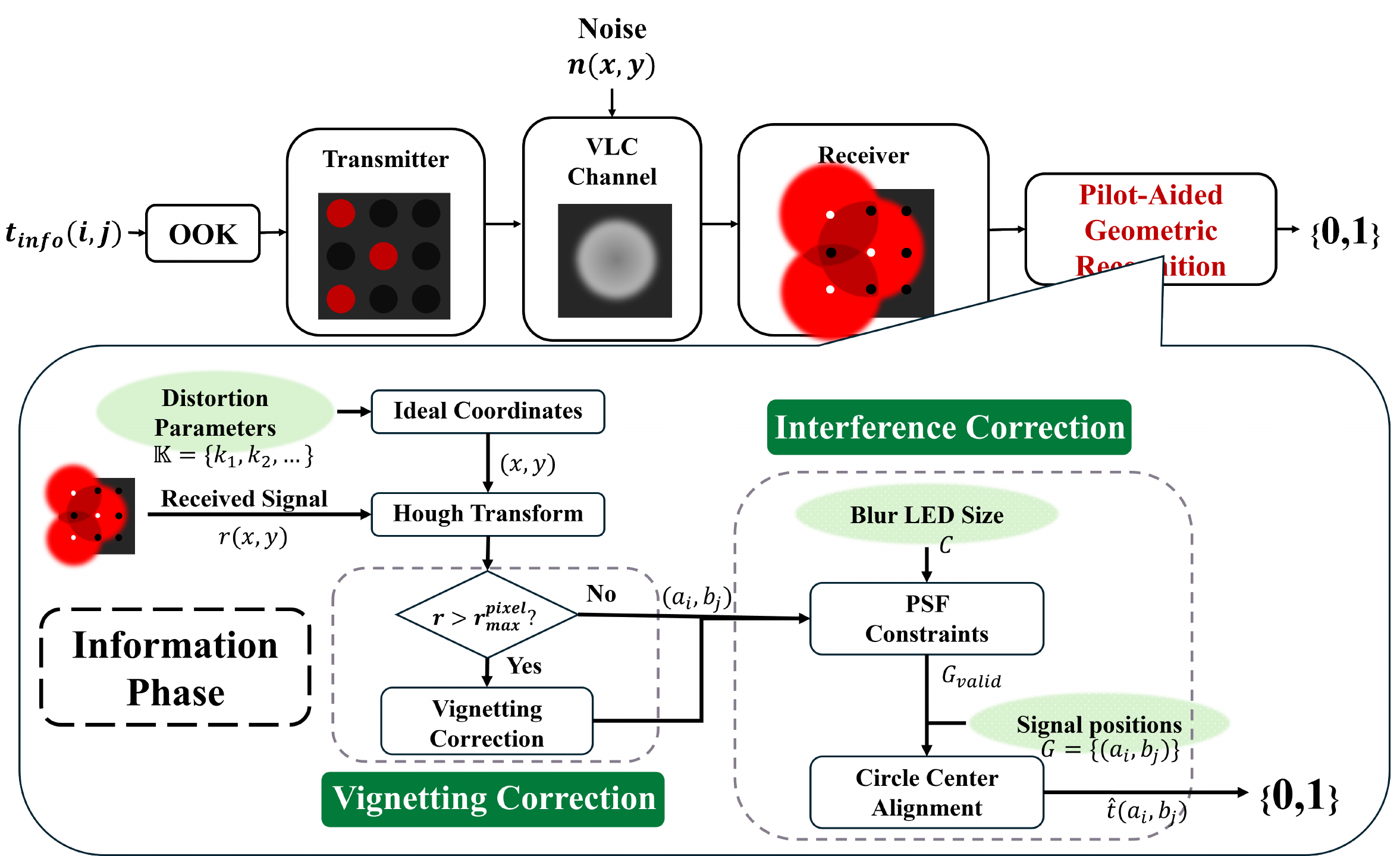} 
\caption{The Method for Detecting Blur LEDs---Information Phase}
\label{fig:Method2}
\end{figure}

Building upon the calibration results from the pilot phase, the second stage—termed the information phase—applies vignetting-aware compensation and interference suppression techniques to extract transmitted signals. As shown in \figurename~\ref{fig:Method2}, this phase refines the decoding by addressing brightness loss and overlapping blur artifacts, ultimately enabling accurate symbol recovery.

\subsection{Information Symbol Definition}
During the information phase, let $t_{\mathrm{info}}(i,j) \in \{0,1\}$ denote the information symbol emitted by the LED located at grid position $(i,j)$. Unlike the pilot phase, the information symbols are not subject to any spatial activation constraint, and adjacent LEDs may be simultaneously active. As a result, the received image may contain severe spatial-domain inter-symbol interference (ISI) caused by overlapping point spread functions (PSFs), in addition to blur, radial distortion, and vignetting effects.

Decoding of the information symbols relies on three parameters obtained from the pilot phase: ideal coordinates(the distortion parameters $\mathbb{K}$), blur LED size ($C_{exp}$), and signal positions ($G$).

\subsection{Ideal Coordinates}
\label{sec:Ideal_Coordinates}
Using the distortion parameters $\mathbb{K}$ defined in eq\eqref{eq.a-b}, 
the ideal (undistorted) coordinates $(a_i, b_j)$ can be recovered.

\subsection{Hough Transform}
To extract the blurred LEDs from the image frame, we apply the circular Hough Transform\cite{Hough}. However, unlike standard applications, we constrain the detection range based on the estimated blur diameter from the pilot image.

Based on the received image $r(x,y)$, circular patterns arising from the PSF of active LEDs are extracted by applying the Hough Transform, resulting in a candidate circle set:
\begin{equation}
  \mathcal{C}_{\text{Candidate}} = \{ (a_k, b_k, R_k) \}_{k=1}^{K} 
  \label{eq:C_candidate}
\end{equation}
where \( (a_k, b_k) \) denotes the center, $k$ is the total number of candidate circles and \( R_k \) is the radius of the \(k\)-th detected circle. Each element in $\mathcal{C}_{\text{Candidate}}$ corresponds to a circular pattern in $r(x,y)$ that may represent an active blur LED signal.

\subsection{Vignetting Correction}
\label{sec:Vignetting Correction}
Given that the candidate circle set $\mathcal{C}_{\text{Candidate}}$, the radial distance $r_{i,j}$, 
and the maximum pixel radius $r_{\max}^{\text{pixel}}$ are known as defined in 
eq\eqref{eq:C_candidate}, eq\eqref{eq:r_ij}, and eq\eqref{eq:r_max}, the following procedure is applied.

To reduce interference caused by vignetting, an additional step based on image detection is applied. The principle is as follows: for each detected circular spot $  \mathcal{C}_{\text{Candidate}} = \{ (a_k, b_k, R_k) \}_{k=1}^{K} $, 

If the distance \(r_{i,j} > r_{\max}^{\text{pixel}}\), the circle is divided into 16 equal angular sectors labeled \( s = 0, 1, \ldots, 15 \), each representing a $22.5^\circ$ segment. To determine the index \( s \) corresponding to the direction \( \vec{v} \), the angle \( \phi_k \) is computed by:
\begin{equation}
\phi_k = \arctan2(b_k - b_c,\; a_k - a_c)
\label{eq:phi_k}
\end{equation}
Then \( \ell \) is obtained by quantizing this angle:
\begin{equation}
\ell= \left\lfloor \frac{\phi_k \bmod 2\pi}{2\pi / 16} \right\rfloor
\end{equation}

To evaluate whether the back (opposite) side of the circle is sufficiently illuminated, the set of opposite sectors is defined as:
\begin{equation}
L_k = \left\{ (\ell + 8 + \delta) \bmod 16 \mid \delta = -2, -1, 0, 1, 2, 3 \right\}
\end{equation}

For each sector \( \ell \in L_k \), the mean brightness \( I_s \) is computed. If the average:
\begin{equation}
\frac{1}{6} \sum_{\ell \in L_k} I_s \geq\text{Threshold}
\end{equation}

then the circle is retained; otherwise, it is discarded.

This directional sector analysis allows robust rejection of partially vignetted circles while preserving valid full or near-full circles.

\subsection{Interference Correction}

Based on the estimated $G$ and $C_{\text{exp}}$, we propose the OOK detection algorithm that is a two-stage detection framework that spatially constrained optical point estimation.

\subsubsection{PSF-Constrained Hough Transform}

We define the set of validated approximate centers as
\begin{equation}
    G_{\text{valid}} = \{ (a_k, b_k) \mid |R_k - C_{\text{exp}}/2| \leq \epsilon \cdot\ C_{\text{exp}} \}
    \label{eq:G_valid}
\end{equation}
where $\epsilon$ is a tolerance factor accounting for lens distortion. This constraint eliminates false positives from noise or smaller artifacts.

\subsubsection{Circle Center Alignment}

The validated approximate centers $G_{\text{valid}}$ detected by eq(\ref{eq:G_valid}) are then aligned with predicted grid point set $G$ generated from the pilot image.
For each detected point $(a_k, b_k)\in G_{valid}$, we compute its distance to all grid points and match it to the nearest $(a_i, b_j) \in G$ satisfying:
\begin{equation}
|a_k - a_i| < \theta\cdot \Delta_a,\quad |b_k - b_j| < \theta\cdot \Delta_b
\end{equation}
where $\theta$ is the matching threshold that is smaller than $0.5$. A detected circle is considered valid only if its positional error falls within a predefined threshold, thereby reducing geometric drift and suppressing outliers.

The output information matrix $\hat{t} \in \mathbb{R}^{16 \times 16}$ is a binary matrix defined as:
\begin{equation}
\label{eq:t}
\hat{t}(a_i,b_j) =
\begin{cases}
1, & \text{if } \exists (a_k, b_k) \in G_{\text{valid}} \text{ such that} \\
   &  |a_k - a_i| < \theta \cdot \Delta_a\text{ and } |b_k - b_j| < \theta \cdot \Delta_b \\
0, & \text{otherwise}
\end{cases}
\end{equation}

Each entry represents whether the LED at grid location $(a_i, b_j)$ is detected as active ("1") or inactive ("0"). 


\textbf{Remark:}
In conventional Hough transform, a single active LED signal may trigger multiple redundant circle detections,
resulting in high false positive rates and increased interference. To address this, we incorporate a PSF-constrained method that ensures each valid signal point yields at most one reliable detection. This is further enhanced by a circle center alignment step, which verifies the geometric consistency of detected circles.

\section{RESULTS AND ANALYSES}

\subsection{Experimental Setup}
\begin{figure}[ht]
     \centering
\includegraphics[width=\linewidth]
     {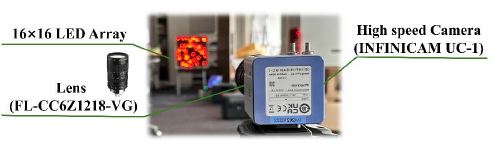}
     \caption{VLC experimental platform view}
     \label{fig:Equipment}
\end{figure}

To validate the proposed decoding framework under realistic optical conditions, we constructed a VLC experimental platform based on a high-density LED array and a high-speed image sensor receiver as illustrated in Fig.~\ref{fig:Equipment}. 
The platform is configured in an indoor environment with adjusting focusing distance $s$ and communication distance $s'$ to simulate interference.
Table~\ref{table:2} summarizes the parameters of VLC experimental platform. 

\begin{table}[h]
\centering
\caption{Parameters of VLC experimental platform}
\label{table:2}
\begin{tabular}{l|r}
\hline
Transmitter & LED Array \\
\hline
LED Blinking Frequency & 500 Hz \\
\hline
LED Array Size & $46 \times 46~(\mathrm{cm}^2)$ \\
\hline
LED Number & $16 \times 16$ \\
\hline
Frame Rate & 1000 fps \\
\hline
$\Delta _{a_{6\text{m}}}$ & 25 pixel\\
\hline
\hline
Receiver & INFINICAM UC-1  \\
\hline
Lens & FL-CC6Z1218-VG   \\
\hline
Pixel Pitch ($P$) & $10~\mu\mathrm{m}$ \\
\hline
Aperture Valure $F$ &1.8\\
\hline
Focal Length $f$ & 30mm \\
\hline
\end{tabular}
\end{table}

\subsection{Inter-Symbol Interference}
To characterize the severity of ISI, we categorize the degree it into three distinct degrees based on the spatial overlap between the blur radius of a transmitting LED and the centers of its neighboring LEDs. Assume $\Delta_a = \Delta_b$, we define the following ISI degrees:\\
\textbf{Degree-1(No interference)}: 
\vspace{-2mm}
$$\Delta_a - \frac{C}{2} > 0$$
The blur circles may partially overlap with adjacent regions, but do not reach the center positions of adjacent LEDs. Signals are well-separated with no interference.\\
\textbf{Degree-2(Moderate Interference)}:
\vspace{-2mm}
$$-\Delta_a < \Delta_a - \frac{C}{2} < 0$$
The blur circle partially overlaps with one neighboring LED center, leading to limited ISI.\\
\textbf{Degree-3 (Severe Interference)}:
\vspace{-2mm}
$$-2\Delta_a < \Delta_a - \frac{C}{2} < -\Delta_a$$ The blur circle overlaps with two adjacent LED centers, causing strong ISI and significant signal contamination.

\begin{figure}[!t]
  \centering
  \begin{minipage}[t]{0.32\linewidth}
    \centering
    \includegraphics[width=\linewidth]{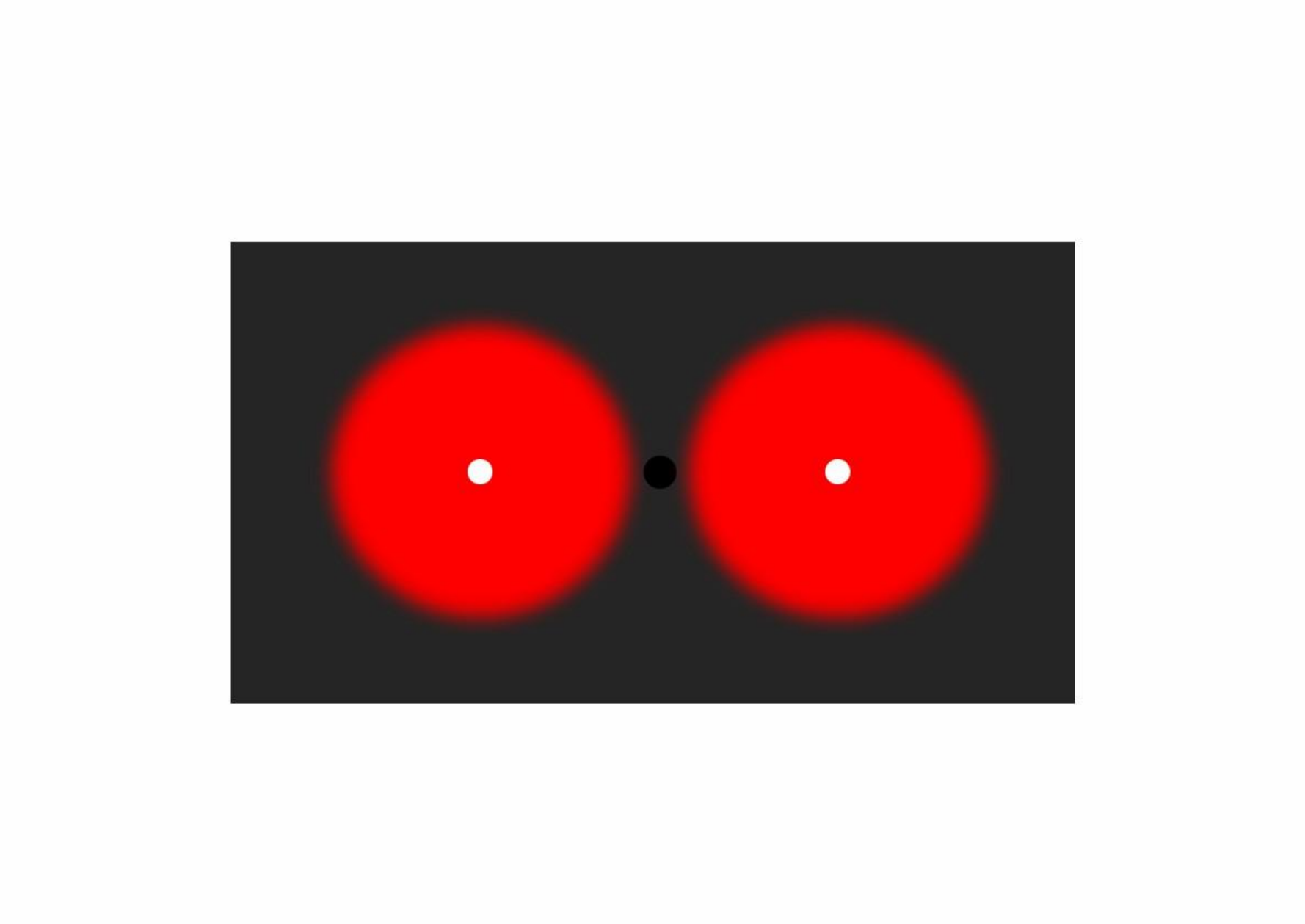}
    \caption*{(a)}
  \end{minipage}
  \begin{minipage}[t]{0.32\linewidth}
    \centering
\includegraphics[width=\linewidth]{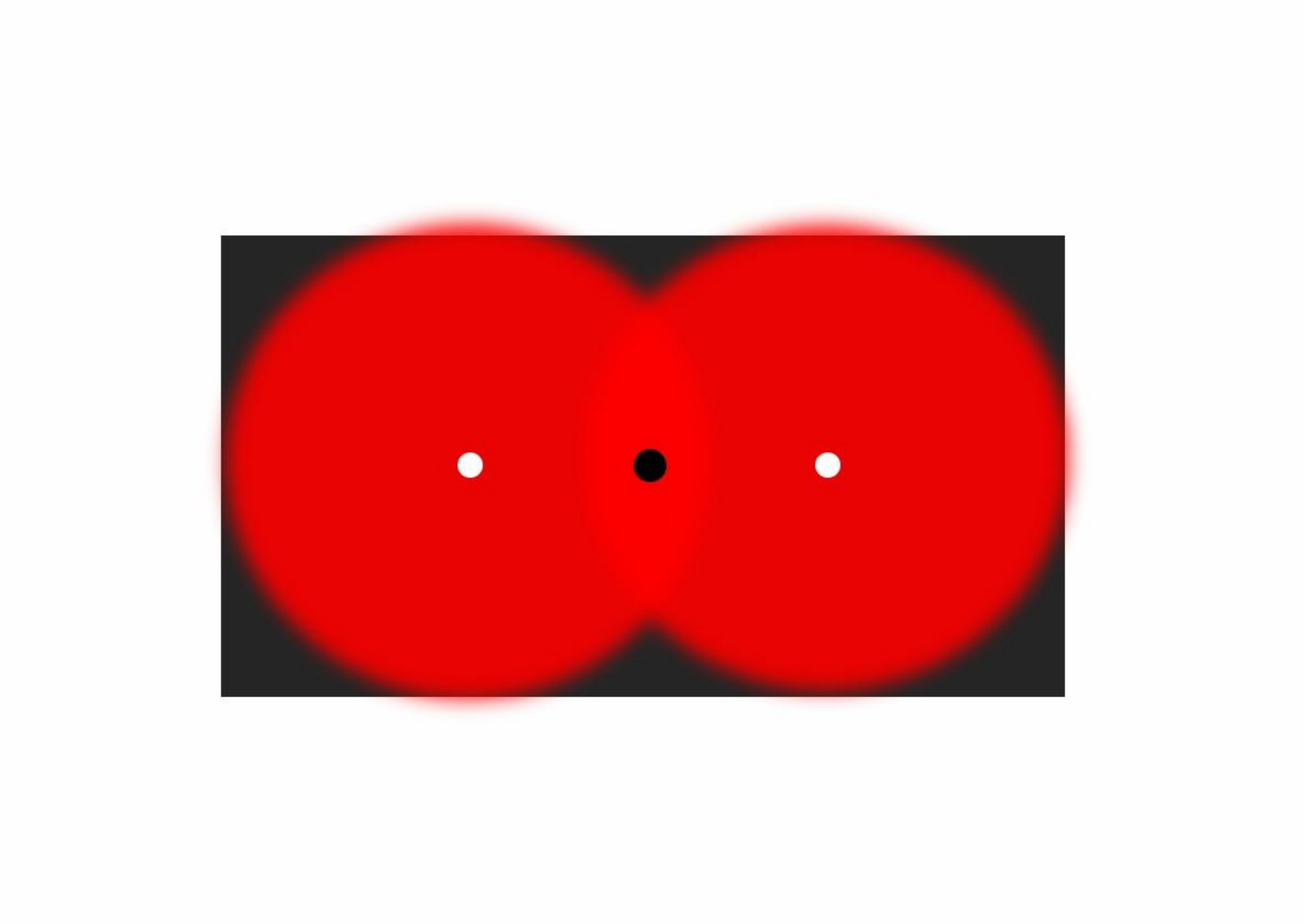}
    \caption*{(b)}
  \end{minipage}
  \begin{minipage}[t]{0.32\linewidth}
    \centering
    \includegraphics[width=\linewidth]{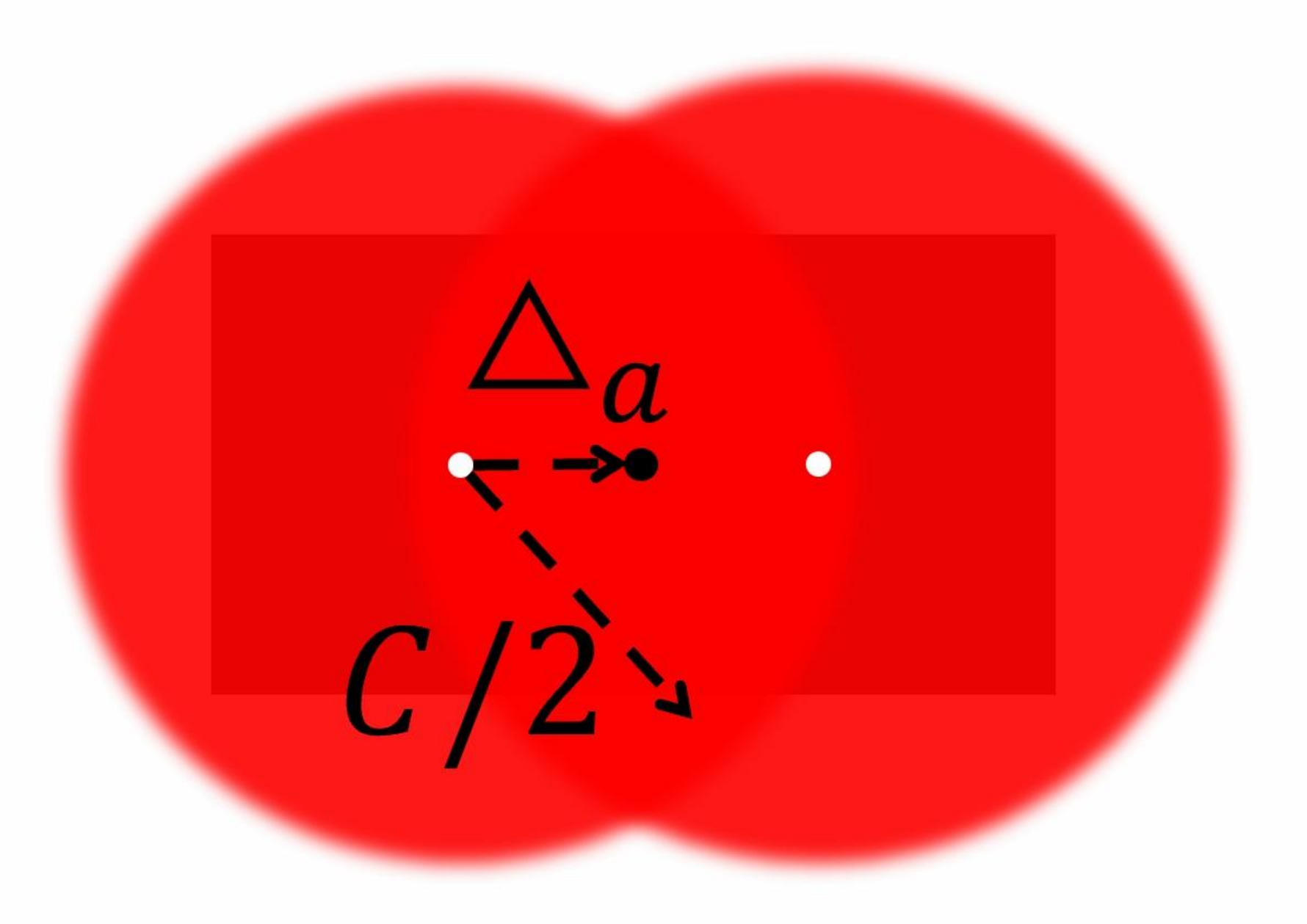}
    \caption*{(c)}
  \end{minipage}
  \caption{Three Degrees of ISI:(a) Degree-1 (b) Degree-2 (c) Degree-3}
  \label{fig:three_images}
\end{figure}

\subsection{Correction on Radial Distortion}

This subsection evaluates the necessity of higher-order radial distortion terms by comparing three models: $k_1$ only, $(k_1,k_2)$, and $(k_1,k_2,k_3)$. The goal is to quantify whether adding higher-order parameters yields meaningful geometric improvement for the pilot-LED grid rectification.

\subsubsection{Estimation of $k_1,k_2,k_3$ in normalized camera coordinates}

Let $(a_i',b_j')$ denote a detected pilot LED coordinate in pixels and
$(c_x,c_y)$ be the camera principal point.
We adopt the definition of the distorted normalized radius in
eq\eqref{eq.r_d}.

Accordingly, the measured normalized radial distance of a detected point is eq\eqref{eq:r_ij}.

The 2D mapping in
eq\eqref{eq:radial_distort} can be equivalently expressed.

In the ideal formulation, the optimal parameters
$\mathbb{K}=\{k_1,k_2,k_3\}$ are obtained by minimizing the full 2D reprojection
error in eq\eqref{eq.a-b}.

 Since $r$ is bounded and typically satisfies $r\in[0,1]$ over most of the image after normalization, the polynomial terms decay rapidly with increasing order ($r^4\ll r^2$ and $r^6\ll r^2$). Therefore, $k_2$ and $k_3$ are expected to be less influential and more weakly observable. 
A practical configuration is:
\\(i) initialize $(k_1,k_2,k_3)$ with $(0,0,0)$, or initialize $k_1$ from a coarse fit and set $(k_2,k_3)=(0,0)$;
\\(ii) optionally apply conservative bounds such as $k_1\in[-10,10]$, $k_2\in[-50,50]$, and $k_3\in[-200,200]$ to prevent unstable extrapolation when the point set is limited. 
In our experiments, warm-started local fitting was sufficient without explicit grid search.

\begin{table*}[t]
\centering
\caption{Pilot information of estimated radial distortion parameters \\and RMSE performance under different model orders (FL-CC6Z1218-VG, $f$=30\,mm).}
\label{tab:RMSE}
\renewcommand{\arraystretch}{1.15}
\begin{tabular}{l c c c c c c}
\toprule
\textbf{Model}
& $k_1$
& $k_2$
& $k_3$
& \textbf{RMSE$_x$} [px]
& \textbf{RMSE$_y$} [px]
& \textbf{RMSE} [px] \\
\midrule

$k_1$
& $-2.8335$ & -- & --
& 0.9768 & 1.0832 & 1.0300 \\

$k_1,k_2$
& $-2.8335$ & $2.1234$ & --
& 0.9707 & 1.0526 & \textbf{1.0116} \\

$k_1,k_2,k_3$
& $-2.8335$ & $2.1234$ & $-4.8678$
& 0.9711 & 1.0547 & 1.0129 \\

\bottomrule
\end{tabular}
\end{table*}

\subsubsection{Why $k_1$ is usually sufficient}
Although $k_2$ and $k_3$ may take non-negligible numerical values, their effective contribution is weighted by higher of $r$.
Specifically, the relative magnitude of the second-order contribution to the first-order one is
\begin{equation}
\frac{|k_2 r^4|}{|k_1 r^2|}=\frac{|k_2|}{|k_1|}r^2,
\end{equation}
and similarly the third-order term scales as $|k_3|r^4/|k_1|$. Since the normalized radius remains bounded ($r\lesssim 1$ in most regions), $r^2$ and $r^4$ suppress higher-order terms strongly. Consequently, $k_2$ and $k_3$ become weakly observable from a limited set of pilot points, and any fitted improvement is often within the corner-localization noise floor.

This behavior is consistent with our ablation study: adding $k_2$ yields only a marginal reduction in straightness RMS and ring variance, while adding $k_3$ provides no further systematic gain. Therefore, the $k_1$-only model achieves the best accuracy--complexity trade-off in typical settings, and higher-order terms are unnecessary unless the field-of-view is extremely wide, edge distortion is exceptionally strong, or a dense and well-distributed calibration grid is available to reliably constrain $(k_2,k_3)$.

We evaluate the geometric rectification performance using the root mean square error (RMSE) between the undistorted grid points and their corresponding ideal grid locations. Specifically, after distortion correction, each recovered point $(\hat{a_i'}, \hat{b_j'})$ is compared with its ideal counterpart $(a_{i}, b_{j})$, and the overall reconstruction error is quantified by the RMSE defined as
\begin{equation}
\mathrm{RMSE}
=
\sqrt{
\frac{1}{36}
\sum_{i=0}^{5}\sum_{j=0}^{5}
\left[
(a_{i}-\hat{a}_i)^2+
(b_{j}-\hat{b}_j)^2
\right]
}.
\end{equation}

The RMSE is evaluated only on the pilot LED positions, which are defined by the sparse pilot grid in eq\eqref{eq:G_pilot}. 
Since the pilot set $G_{\text{pilot}}$ forms a $6 \times 6$ subgrid, the RMSE is computed over a total of $36$ pilot points.

To further analyze the rectification accuracy along different spatial directions, the horizontal and vertical RMSE components are separately defined as
\begin{equation}
\begin{aligned}
\mathrm{RMSE}_x &=
\sqrt{
\frac{1}{36}
\sum_{i=0}^{5}\sum_{j=0}^{5}
(a_{i}-\hat{a}_i)^2},\\
\mathrm{RMSE}_y &=
\sqrt{
\frac{1}{36}
\sum_{i=0}^{5}\sum_{j=0}^{5}
(b_{j}-\hat{b}_j)^2}.
\end{aligned}
\end{equation}

The RMSE directly quantifies the residual geometric deviation of the rectified grid with respect to the ideal lattice, providing a physically interpretable and geometry-consistent measure of distortion correction quality. A smaller RMSE indicates more accurate distortion compensation and improved grid regularity.

As summarized in Table~\ref{tab:RMSE}, although higher-order radial distortion parameters $(k_2, k_3)$ take non-negligible numerical values after optimization, their inclusion yields only marginal improvement in RMSE. In most cases, the first-order model using only $k_1$ already achieves near-optimal geometric rectification accuracy, indicating that higher-order terms contribute limited additional benefit under the considered imaging conditions.

\subsection{Estimation of Diameter of the Blurred LED }

To evaluate the accuracy of blur circle estimation, we compare the experienced blur diameter $C_{\text{exp}}$ with the actually measured blur diameter $C_{\text{Data}}$, where $C_{\text{Data}}$ is the apparent diameter of the blur spot formed by a single active LED directly measured from the captured pilot image through PSF-constrained Hough transform. 

The relative error is computed as: 
\begin{equation} 
\text{Error (\%)} =  \frac{C_{\text{Data}} - C_{\text{exp}}}{C_{\text{Data}}} 
\end{equation}

TABLE~\ref{table:3} summarizes the relative error between the predicted and observed blur diameters on various focusing distances $s$.
Experimental results show that the relative error remains within 10\%. This demonstrates the predictive accuracy and robustness of the proposed PSF-based blur model under real-world optical conditions.

\begin{table}[h]
    \centering
    \caption{The diameter of the blur LED Estimation}
    \label{table:3}
    \renewcommand{\arraystretch}{1.0} 
    \setlength{\tabcolsep}{3pt} 
    \scalebox{0.85}{ 
    \begin{tabular}{cccccccc}
        \toprule
        ISI Degree& $\Delta_a$ & $s' \text{ (m)}$ &  $s \text{ (m)}$ &  $C_{\text{calc}} \text{ (\textmu m)}$ & $C_{\text{exp}} \text{ (px)}$ & $C_{\text{Data}} \text{ (px)}$ & $\text{Error} (\%)$ \\
        \midrule
        1& 33 & 4 &  1.0  & 356  & 43  & 42  & ~2 \\
        2& 22 & 6  & 1.0  & 419  & 54  & 52  & -4 \\
        1& 22 & 6  & 1.2  & 335  & 44  & 42  & -5  \\
        3& 15 & 8  & 1.0  & 461  & 60  & 62  & ~3   \\
        2& 15 & 8  & 1.2  & 355  & 44  & 44  & ~0   \\
        1& 15 & 8  & 1.5  & 271  & 30  & 32  & ~6   \\
        \bottomrule
    \end{tabular}
   }
    \label{tab:error_calculation}
\end{table}

\subsection{PSF Constraints and Circle Center Alignment}

\begin{figure}[!t]
     \centering
     \includegraphics[width=0.8\linewidth]
     {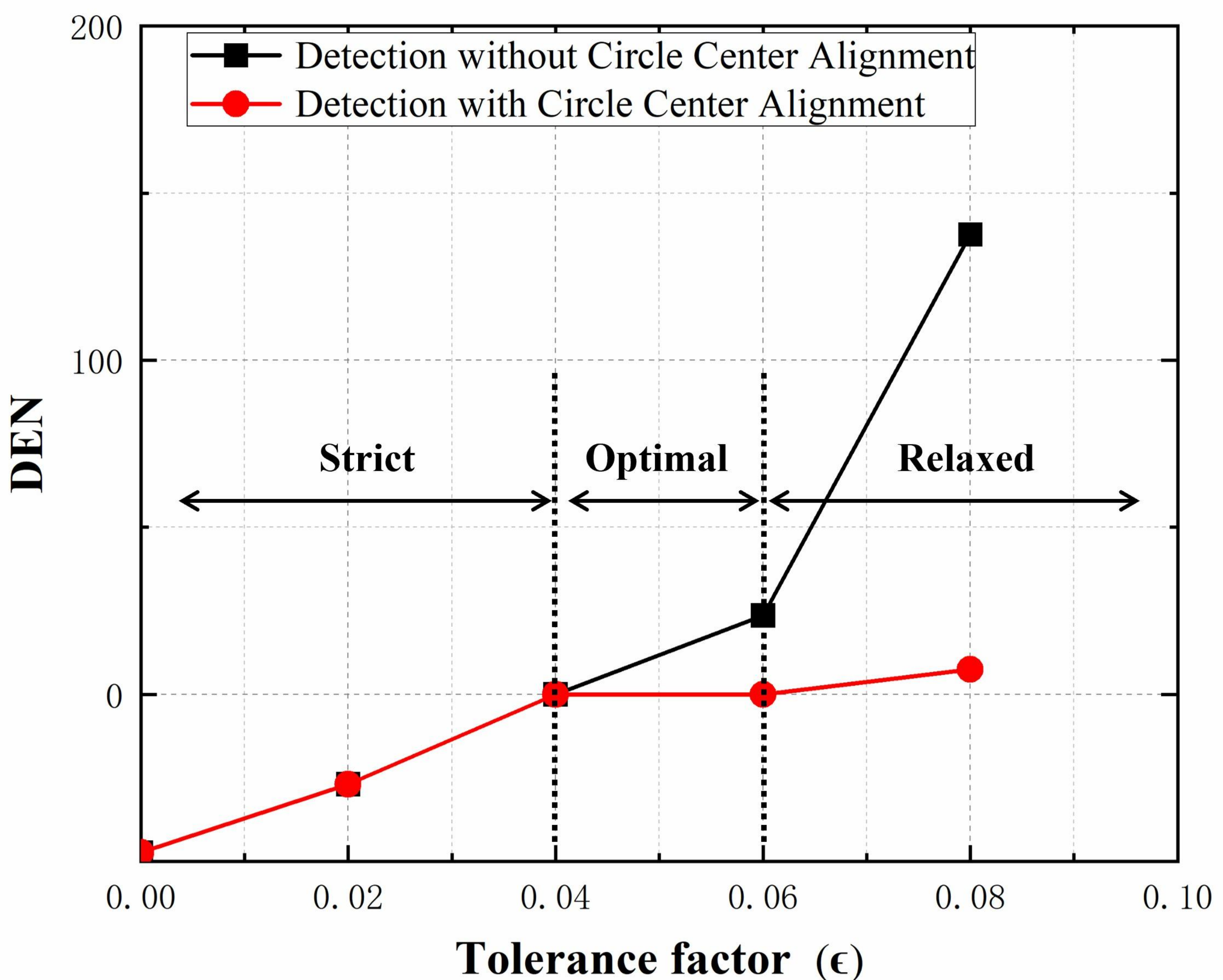}
     \caption{Detection with ISI under PSF Constraints}
     \label{fig:PSA-HTCCA}
\end{figure}

\figurename~\ref{fig:PSA-HTCCA} shows detection accuracy of the proposed method with variable tolerance factor $\epsilon$.
We consider a setup where the object distance is $s'=6$m and the focusing distance is $s=1$m, resulting in significant optical defocus under ISI Degree-2.
A total of 80 LEDs are activated, corresponding to approximately 30\% of the full LED array. 
To control the degree of ISI, the layout is manually designed to ensure that no LED blur spot overlaps with more than five others.

Detection accuracy is evaluated using the Detection Error Number (DEN), defined as: 
\begin{equation}
     \text{DEN}= {n_{\text{detected}} - n_{\text{actual}}} 
\end{equation}
where $n_{\text{actual}}$ denotes the number of “1” symbols in the transmitted signal and $ n_{\text{detected}}$ is the number of detected “1” symbols. For comparison, we also include the DEN of the PSF-constrained circle detection method without the proposed circle center alignment.

As shown in \figurename~\ref{fig:PSA-HTCCA}, when $\epsilon < 0.04$, DEN is negative, indicating under-detection, i.e., some active LEDs are misdetected. This is because the stricter tolerance leads to tighter matching, which causes some valid circles to be rejected.

As $\epsilon$ increases, DEN becomes positive. In this case, both methods tend to misdetect "0" signal to "1" due to over-detected. However, the method without circle center alignment suffers from significantly more false positives, as relaxed tolerance leads to excessive circle detection.

\subsection{Maximum Allowable Chief Ray Angle Estimation}
To estimate which areas on the image sensor
plane can be fully illuminated without being blocked by the lens
barrel or aperture stop. We analyze the geometric relationship between lens structure and incident light angles. As an example, we consider a Ricoh FL-CC6Z1218-VG lens (focal length $f= 30 mm$) used with the INFINICAM UC-1 camera (pixel size  $P=10$ µm) and its lens length \( L = 90 \,\mathrm{mm} \), the maximum chief ray angles and corresponding effective image radii for various F-numbers are summarized as Table~\ref{Chief Ray Angle} showed:

\begin{table}[h]
\centering
\caption{Chief Ray Angle and Effective Image Radius under Various F-numbers}
\label{Chief Ray Angle}
\begin{tabular}{ccccc}
\toprule
F  &   D & \( \theta_{\text{max}} \) &  $r_{\text{max}}$& $r_{\text{max}}^{\text{pixel}} $ \\
\midrule
1.8  &16.7& 5.3$^\circ$ & 2.78 & 278 \\
4 & 7.5 &2.4$^\circ$ & 1.26 & 126 \\
8 & 3.75 &1.19$^\circ$  & 0.62 & 62 \\
\bottomrule
\end{tabular}
\end{table}

As shown, larger apertures (lower F-numbers) permit greater chief ray angles and allow wider image coverage. In contrast, smaller apertures (higher F-numbers) reduce the usable field and increase the likelihood of mechanical vignetting at the image periphery. For the INFINICAM UC-1.1, which has an effective sensor size of 12.8 mm $\times$ 10.24 mm (diagonal $\approx$ 16.3 mm), even at F=1.8 the image circle does not fully cover the entire sensor area without vignetting.

\begin{table}[h]
\centering
\caption{Visible area thresholds beyond the effective radius ($R=30$ px)}
\label{tab:Chief Ray}
\begin{tabular}{c m{0.08\textwidth} c c}
\toprule
Level & Figure & Visible Area Ratio & $r_{\text{max}}^{\text{pixel}}$ \\
\midrule
Level 0 & \centering\includegraphics[width=0.03\textwidth]{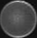} & $100\%$        & $<278$        \\
Level 1 & \centering\includegraphics[width=0.03\textwidth]{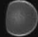} & $100\%\sim75\%$ & $278\sim290$ \\
Level 2 & \centering\includegraphics[width=0.03\textwidth]{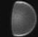} & $75\%\sim50\%$  & $290\sim308$ \\
Level 3 & \centering\includegraphics[width=0.03\textwidth]{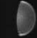} & $<50\%$        & $>308$        \\
\bottomrule
\end{tabular}
\end{table}

\begin{figure}[!t]
     \centering
\includegraphics[width=\linewidth]
     {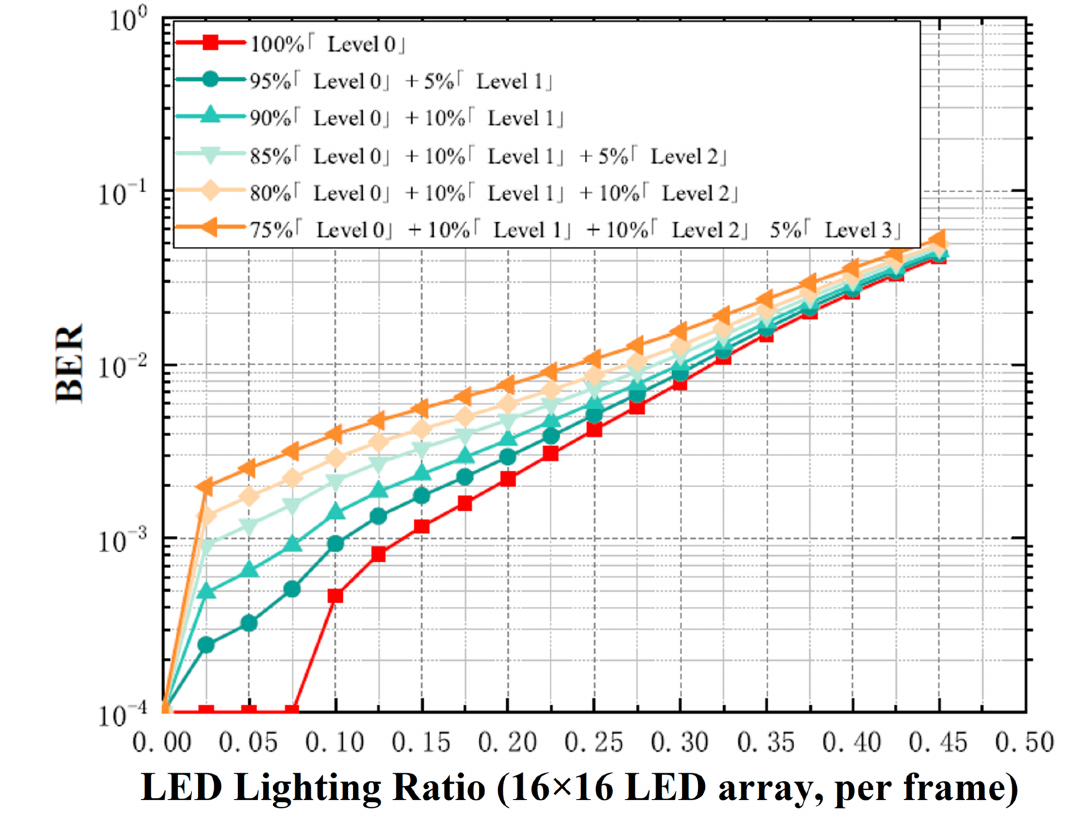}
     \caption{BER degradation trend under varying visible area ratios}
     \label{fig:Chief Ray}
\end{figure}

Under the experimental condition of $F=1.8$, $s'=6\,\text{m}$, and $s=1\,\text{m}$, Table~\ref{tab:Chief Ray} and Fig.~\ref{fig:Chief Ray} jointly reveal the quantitative relationship between the visible area ratio and BER performance. For Level~0 (100\% visible area without vignetting), among 16$\times$16 = 256 signal points, when the lighting ratio is 0.15, the system maintains a BER lower than $10^{-3}$, indicating stable and reliable decoding in the central region.  

As the LED array approaches the sensor edge and Level~1 vignetting occurs, the performance begins to degrade. Under the condition of 5\% Level~1, when the lighting ratio is 0.1, the BER still remains below $10^{-3}$. Even with 10\% Level~1, when the lighting ratio decreases to 0.075, the BER is still maintained under $10^{-3}$. 

However, when further decreasing to Level~2, maintaining a BER of $10^{-3}$ requires the lighting ratio to be constrained within 0.025–0.05. Once Level~3 vignetting appears, the error probability rises sharply, and the system loses its reliability.  

It is important to note that the occurrence of vignetting is progressive: Level~1 must first appear, followed by Level~2, and finally Level~3. The “10\% Level~1 condition” mentioned here is specific to the current experimental optical setup, and the exact proportion may vary under different lens parameters and imaging geometries.  

\begin{figure}[!t]
     \centering
     \includegraphics[width=\linewidth]
     {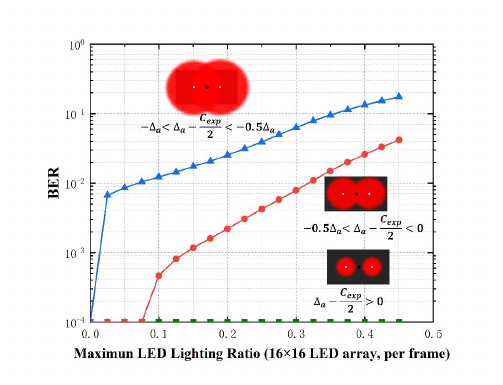}
     \caption{Bit Error Rate under varying Maximun LED lighting ratio}
     \label{fig:BER}
\end{figure}

\subsection{BER and Throughput}
\figurename~\ref{fig:BER} evaluate communication reliability under varying LED lighting ratios for the\textit{ \textit{three degrees}} of ISI. 
The BER is the proportion of incorrectly decoded bits relative to the transmitted bits in each frame as $|\text{DEN}|/n_{\text{acutal}}$.

The\textit{ LED lighting ratio} is the proportion of lit LEDs to the total number of LEDs in the array during each frame, expressed as $L/N^2$, where $L$ is the number of LEDs that are lit on in a frame over $N^2$ total number of LEDs array.
In the experience, we set $\epsilon=0.04$. 

When ISI is at Degree-1, the spatial distance is preserved, and the BER remains extremely low (below $10^{-5}$). Under moderate ISI (Degree-2), BER remains low at smaller lighting ratio but increases at the lighting ratio rises. This is because five or more blur circles may overlap in a given region, resulting edge patterns cluttered and irregular. In the case of severe ISI (Degree-3), multiple surrounding circles introduces additional interference, and the superimposed light intensity can exceed the camera's saturation threshold. As a result, the boundaries of the resulting blur circles disappear, making them indistinguishable.

To mitigate BER under high interference conditions, techniques such as error-correcting codes, constrained modulation schemes, and dynamic PSF-aware decoding strategies can be employed to futher reduce the impact of multi-LED overlap and adapt to varying blur conditions.

\begin{figure}[!t]
     \centering
    \includegraphics[width=\linewidth]{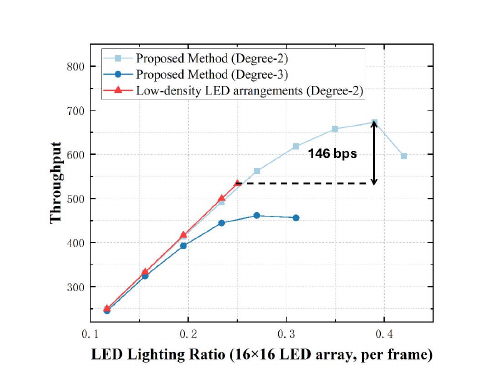}
     \caption{Throughput (Proposed Method vs. Other Method)}
     \label{fig:Throughput}
\end{figure}

\figurename~\ref{fig:Throughput} compares the throughput performance of the proposed high-density system at degree-2 and 3 with a conventional low-density configuration \cite{lowdensity} under various LED lighting Ratio. The throughput is defined as the number of correctly decoded bits per minute.

The conventional low-density setup \cite{lowdensity} arranges the LEDs sparsely to avoid spot overlap at center positions, which effectively eliminates ISI. However, this interference-free layout comes at the cost of a significantly reduced data rate, approximately one-fourth of that of a full-density congiguration (our method). In this scheme, signal detection is based solely on light intensity, where the presence of a light spot is interpreted as a binary "1" signal, and its absence as a "0" signal. Under moderate ISI conditions (Degree-2), the low-density LED setup achieves a BER of zero, and its throughput is comparable to that of our proposed method when lighting ratio is smaller than 0.25. 
However, under severe interference (Degree-3), this approach fails completely due to extensive spot overlap, which renders signal boundaries indistinguishable and causes decoding to collapse.

In contrast, under degree-2 interference, the proposed method achieves a maximum data rate of approximately 0.38, \textbf{yielding around a 25\% improvement} compared with the low-density baseline. More importantly, it remains decodable even under severe degree-3 interference.

As shown in \figurename~\ref{fig:BER} and \figurename~\ref{fig:Throughput}, the proposed method achieves higher throughput under Degree-2 interference as the lighting ratio increases. Although BER increases under Degree-3 conditions, the system maintains decodability at lower lighting ratios. These results confirm that the proposed framework provides significant improvements in both transmission efficiency and decoding reliability across varying levels of spatial interference.

\section{Conclusion}
This paper proposed a robust decoding framework for high-density LED array VLC systems under severe spatial-domain inter-symbol interference (ISI). By leveraging pilot-guided geometric recognition and a PSF-constrained Hough transform with circle center alignment, the proposed method accurately detects and decodes overlapping blurred LED spots. Experimental results under varying ISI degrees demonstrate that our method achieves low BER and high throughput even in challenging optical conditions. Compared with conventional low-density configurations, it maintains decoding reliability while significantly improving the transmission rate. Furthermore, by integrating radial distortion correction and vignetting-aware validation, the framework effectively compensates for geometric deformation and peripheral spot degradation, ensuring stable decoding performance across the entire field of view. This work offers a promising solution for scalable VLC systems and can be extended to future scenarios such as vehicular signaling and dense indoor IoT deployments.

\bibliographystyle{IEEEtran}
\bibliography{sample}
\end{document}